\documentclass[%
reprint,
secnumarabic,
amssymb, amsmath,
aps,cha,
groupedaddress,
frontmatterverbose,
]{revtex4-1}
\usepackage{graphicx, setspace}
\usepackage[usenames, dvipsnames]{color}
\usepackage{graphicx}
\usepackage{caption}
\usepackage{subcaption}
\usepackage[colorlinks=true,linkcolor=blue]{hyperref}
\expandafter\ifx\csname package@font\endcsname\relax\else
\expandafter\expandafter
\expandafter\usepackage
\expandafter\expandafter
\expandafter{\csname package@font\endcsname}%
\fi
\begin{document}
\title{Breaking of Large Amplitude Relativistically Intense Electron Plasma Waves in a Warm Plasma}%
\author{Arghya Mukherjee$^{1, 2}$}
\email{arghya@ipr.res.in}
\author{Sudip Sengupta$^{1, 2}$}
\affiliation{$^1$Institute for Plasma Research, Bhat, Gandhinagar 382 428, India, \\
$^2$Homi Bhabha National Institute, Training School Complex, Anushakti Nagar, Mumbai 400085, India}
%
%
\begin{abstract}
In this paper, the effect of finite electron temperature on the space-time evolution and breaking of a large amplitude relativistically intense electron plasma wave has been studied, using a 1-D relativistic Particle-in-Cell (PIC) code. We have found that for phase velocities for which $\gamma _\phi \ll 1 + \frac{k_BT_e}{mc^2}$, the wave damps within a few plasma period and essentially follows the relativistic Landau Damping rate predicted by Buti \cite{buti}. In the opposite regime (\textit{i.e.} for $\gamma _\phi \gg 1 + \frac{k_BT_e}{mc^2}$) we have observed that waves propagate through the system for a long period of time and in small amplitude limit follow the relativistic warm plasma dispersion relation \cite{buti, misra, pegoraro_porcelli, remi, melrose}. Further we have  demonstrated that in the same regime (\textit{i.e.} for $\gamma _\phi \gg 1 + \frac{k_BT_e}{mc^2}$), for the phase velocities less than the velocity of light $c$, like the cold plasma  Akhiezer - Polovin wave \cite{prabal_main}, in a warm plasma also, relativistically intense waves break via phase mixing when perturbed by an arbitrarily small amplitude longitudinal perturbation. Using the simulation results, we have also shown that the phase mixing time scale in a warm plasma can be interpreted using Dawson's formula \cite{dawson_main} for phase mixing time for a non-relativistic cold inhomogeneous plasma, which is based on out of phase motion of neighbouring oscillators constituting the wave.
\end{abstract}

%
%
%
\maketitle
%
%
\section{Introduction}
The breaking of large amplitude electron plasma waves/oscillations has been receiving a great deal of attention since 1959 due to it's basic nature and practical importance \cite{dav_book, kruer_book, gibbon, mulsar_book}. It has wide applications to some current research problems ranging from laboratory plasma to astrophysical plasma where breaking of large amplitude relativistically intense electron plasma waves are routinely encountered \cite{Tajima_79, Umstadter_01, esarey_IEEE, esarey_rev_mod, Sprangle_90, mori_94, joshi_81, botha, Faure_04, Mangles_04, Geddes_04, Modena_96, Malka_08, marques_1, marques_2, sayak_2,  sayak}. For example, recent experiments on plasma acceleration by laser and particle beam have shown that the breaking of excited plasma oscillations/waves plays a major role in the particle acceleration process \cite{Tajima_79, Umstadter_01, esarey_IEEE, esarey_rev_mod, Sprangle_90, mori_94, joshi_81, botha, Faure_04, Mangles_04, Geddes_04, Modena_96, Malka_08}. Wave breaking is also important for first ignition concept in inertial confinement thermonuclear fusion \cite{Tabak_94, Honda_96, ss_nf}. The concept of wave breaking in a cold homogeneous plasma was introduced by Dawson \cite{dawson_main}, where thermal motion was neglected and ions were fixed. Dawson demonstrated that the amplitude of applied  perturbation can not be increased beyond a critical limit, known as wave breaking limit, as the trajectory of the neighbouring electrons constituting the oscillation/wave start to cross each other beyond this limit. This results in fine scale mixing of various parts of the oscillation which destroys the oscillation/wave. But, when non-linear density perturbations are excited in a large amplitude plasma wave, thermal effects may become important as the electron thermal pressure may not allow the density compression to build up as predicted by the simple cold plasma fluid model. In 1971, Coffey \cite{coffey_wb} investigated this phenomena for electron plasma wave in a warm plasma by using the simplest distribution \textit{i.e.} ``water-bag'' distribution \cite{wb_68} for electrons. Unlike in the cold plasma case where the wave-breaking limit is defined by trajectory crossing, in the case of warm plasma Coffey defined wave breaking as the trapping of background plasma electrons in the wave potential. An analytical expression for the maximum electric field amplitude and density amplitude as a function of the electron temperature has been derived which shows that temperature effects significantly reduces the wave breaking limit \cite{coffey_wb}. Unlike the nonrelativistic warm plasma case, where Coffey's limit is the one and only existing theoretical wave breaking limit available in the literature (till date), the relativistic counterpart contains several theoretical results given by several group of authors in last three decades. These are as follows:  

In 1988, Katsouleas and Mori \cite{Katsouleas_88, scripta}, first extended the calculations carried out by Coffey \cite{coffey_wb} by including relativistic mass variation effects. By using a relativistic water bag model, an analytical expression for the maximum electric field amplitude ($E_{KM}$) that can be sustained by a relativistically intense electron plasma wave in a warm plasma has been derived as a function of electron temperature and Lorentz factor ($\gamma _\phi$), which can be written as
\begin{equation}
\frac{eE_{KM}}{m\omega_pc} = \lambda ^{-1/4}\left[ ln(2\gamma _\phi ^{1/2} \beta ^{1/4}) \right]^{1/2}  \label{eqaa}
\end{equation}
where $\lambda = 3k_BT_e/mc^2$ is the normalised electron temperature.
 The authors \cite{Katsouleas_88, scripta} strictly mentioned that the above expression [Eq.(\ref{eqaa})] is valid only in the ultrarelativistic regime which is defined as $\gamma _\phi ^2\lambda >> 1$. In the same year Rosenzweig \cite{rosenzweig_pra_88} presented another expression of maximum electric field amplitude ($E _{ROS}$, in the limit $v_\phi \rightarrow c$) as a function of electron temperature. The analytical expression for $E_{ROS}$ is given by 
\begin{equation}
\frac{eE_{ROS}}{m\omega_pc} = \left[ \frac{4}{9\lambda} \right]^{1/4}       \label{eqbb}
\end{equation}
Similar wave breaking limit [same as Eq.(\ref{eqbb})] was obtained by Sheng and Meyer-ter-Vehn \cite{meyer} in 1997, using a different set of equations \cite{newcomb_82, newcomb_2, siambis_87}. Recently Schreoder \textit{et. al.} \cite{schroeder_rapid, schroeder_06_POP} proposed a new model of relativistic warm fluid theory and derived the following two expressions for wave breaking amplitude ($E _{SES}$) in the limits $\gamma _\phi ^2\lambda >> 1$ and $\gamma _\phi ^2\lambda << 1$ (laser wake field regime) respectively. These expressions respectively can be written as
\begin{equation}
\left[ \frac{eE_{SES}}{m\omega_pc} \right]^2 = \left( \frac{2}{3}\right)^{3/2} \left(\frac{\lambda}{3}\right)^{-1/2}\left[ 1 - \left\lbrace \frac{\lambda}{2}\right\rbrace ^{1/2} \right] ^3  \label{eqcc}
\end{equation}
\begin{equation}
\left[ \frac{eE_{SES}}{m\omega_pc} \right]^2 = 2(\gamma _\phi -1) - \gamma _\phi \left[ \frac{4}{3}\left( \gamma _\phi ^2\lambda \right) ^{1/4} - \left( \gamma _\phi ^2\lambda \right) ^{1/2} \right]  \label{eqdd}
\end{equation}
 Later Trines \textit{et. al.} \cite{trines_2006} extended the calculations of Katsouleas $\&$ Mori \cite{Katsouleas_88, scripta} to the regime $\gamma _\phi ^2\lambda << 1$ and derived the following expression of wave breaking limit
\begin{equation}
\left[ \frac{eE_{TN}}{m\omega_pc} \right]^2 = 2(\gamma _\phi -1) - 2\gamma _\phi \left[\left( \gamma _\phi ^2\lambda \right) ^{1/4} - \left( \gamma _\phi ^2\lambda \right) ^{1/2} \right]  \label{eqee}
\end{equation}
 All these theoretical results \cite{Katsouleas_88, rosenzweig_pra_88, scripta, meyer, schroeder_rapid, schroeder_06_POP, trines_2006} clearly indicate that, thermal effects significantly reduces the wave breaking limit from the cold plasma Akhiezer - Polovin limit \cite{AP} ($eE _{AP}/m\omega_pc = \sqrt{\gamma _\phi -1}$, derived by Akhiezer and Polovin \cite{AP} in 1956 for a travelling wave in a relativistic cold plasma). Physically it is expected, because the tendency of plasma density to increase to infinity at the breaking point is opposed by the thermal pressure term and the inclusion of thermal velocity of the particles in the direction of wave propagation enables them to get trapped at a lower amplitude of the wave.

\begin{figure}[htbp]
\begin{center}
\includegraphics[scale=0.4]{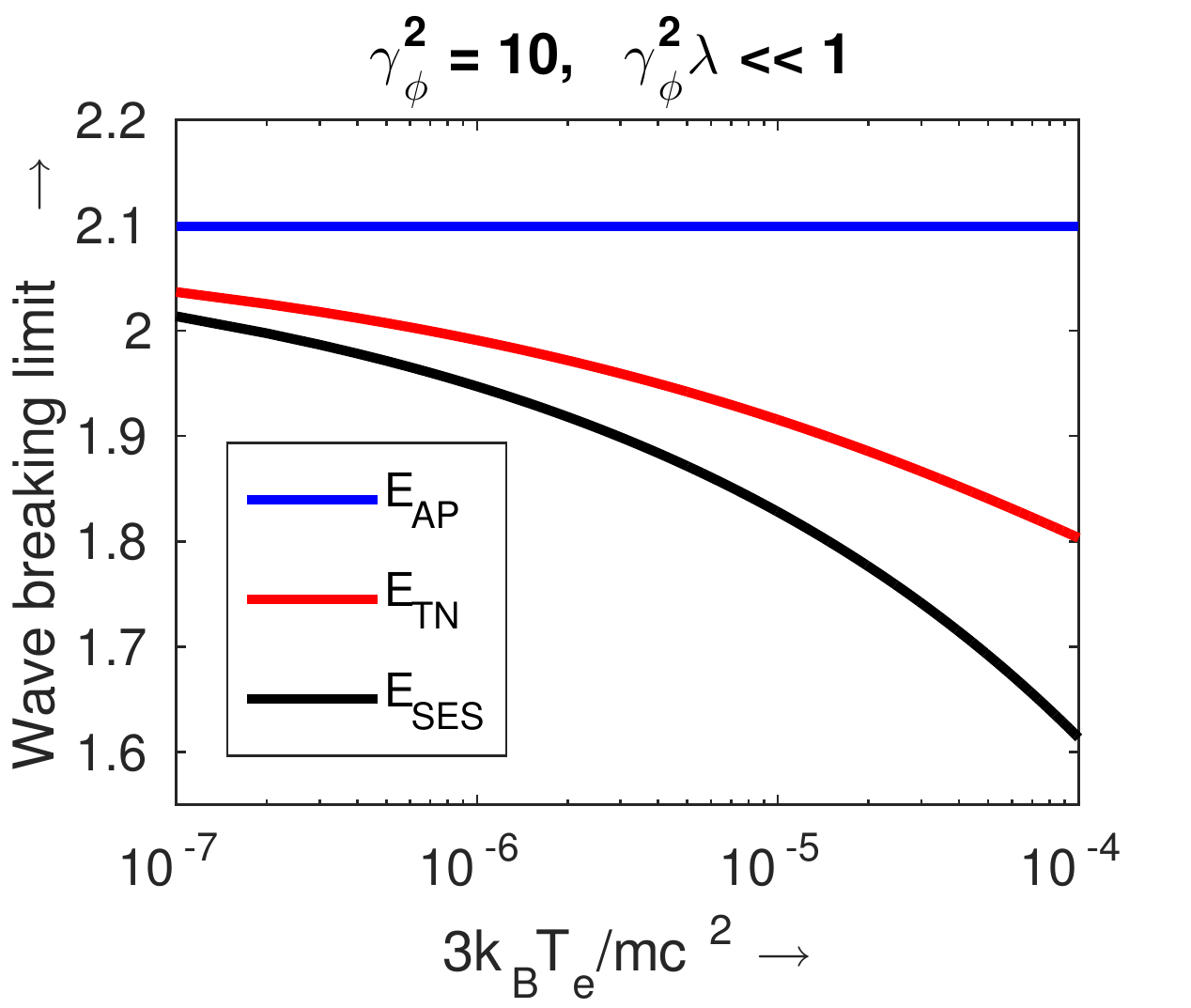}
\caption{Wave breaking limit as a function of $\lambda$ in laser wake field regime}
\label{fig5a}
\end{center}
\end{figure}
\begin{figure}[htbp]
\begin{center}
\includegraphics[scale=0.4]{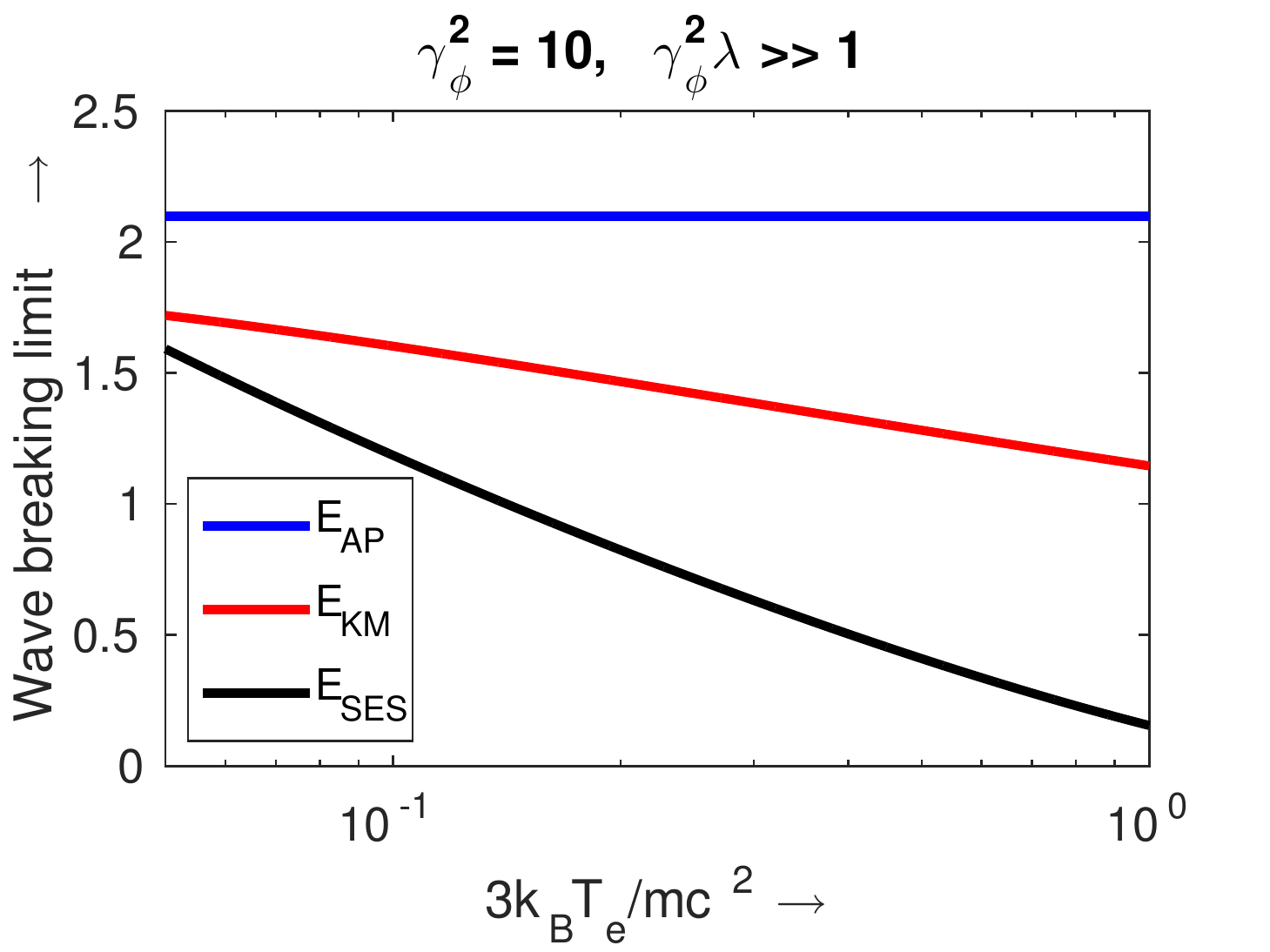}
\caption{Wave breaking limit as a function of $\lambda$ in ultrarelativistic regime}
\label{fig5b}
\end{center}
\end{figure}

  It should be noted here that, considerable amount of work has been contributed by several authors \cite{Katsouleas_88, rosenzweig_pra_88, scripta, meyer, schroeder_rapid, schroeder_06_POP, trines_2006} to this subject over past three decades, mainly focusing on the theoretical analysis by assuming wave like solutions of relativistic Vlasov - Maxwell's equation and these results sometimes lead to different conclusions \cite{trines_2006, attack_on_trines, by_trines}. As for example, in Figs.(\ref{fig5a}) and (\ref{fig5b}) we have shown the variation of wave breaking limits derived by different authors (as discussed above) as a function of electron temperature $\lambda (= 3k_BT_e/mc^2)$ for a fixed value of Lorentz factor ($\gamma _\phi ^2 = 10$). These figures clearly show differences in results obtained by different authors \cite{Katsouleas_88, scripta, schroeder_rapid, schroeder_06_POP, trines_2006} even in the same parameter domain. The only similarity is that all the expressions in the regime $\gamma _\phi ^2\lambda \ll 1$ approach the cold plasma Akhiezer - Polovin \cite{AP, gibbon} limit ($E_{AP}$) in the limit $\lambda \rightarrow 0$. Here we would like to mention that although Trines \textit{et. al.} \cite{trines_2006, attack_on_trines, by_trines} have made some attempt to resolve the above differences by giving mathematical arguments which are essentially based on Taub's inequality \cite{taub} and closure of the hierarchy of the relativistic fluid equations, but to the best of our knowledge, till date there is no consensus on a suitable theoretical model/expression for studying the breaking of relativistically intense electron plasma waves in a thermal plasma. So at this present situation, it is imperative to conduct a numerical experiment on the space-time evolution and breaking of a large amplitude relativistically intense plasma wave in a warm plasma with a relativistically correct velocity distribution. Thus, in this paper we carry out Particle-in-Cell (PIC) simulations in order to investigate the effect of electron temperature on the breaking of a relativistically intense electron plasma wave in a warm plasma where electron's velocity distribution is a J\'{u}ttner \cite{synge}. 
  
  Our first aim would be to excite a travelling wave with relativistic speed in a J\'{u}ttner -Synge \cite{synge} plasma which propagates without damping for a large period of time and next to check its sensitivity towards a small amplitude longitudinal perturbation. Because it is also very crucial to check whether the above limits hold in the presence of a small amplitude perturbations or do they phase mix \cite{sudip_prl, Sengupta_09, Sengupta_11, sudip_conference, prabal_main} like cold plasma Akhiezer - Polovin wave; as in a realistic experiment, there will always be some noises associated with the excited wave. From the present understanding it is expected that due to the applied perturbation, the characteristic frequency would acquire a spatial dependency which would lead to phase mixing \cite{sudip_prl, Sengupta_09, Sengupta_11, sudip_conference, prabal_main}. Therefore, we measure the characteristics frequency of the wave at each position in space, for both the cases without and with the external perturbation.

Thus in order to reach our goal, in section \ref{c5section2}, we first load Akhiezer - Polovin type initial conditions (parametrized by amplitude $u_m$ and phase velocity $\beta _{\phi}$) in our PIC code. Along with this, we also load a finite electron temperature (J\'{u}ttner - Synge distribution) to the background. Here we note that, the inclusion of non-zero electron temperature would try to damp the excited wave within a few plasma period by relativistic Landau damping effect - which would swamp out the wave breaking physics. This damping rate crucially depends on the phase velocity of the wave and the background electron temperature. It is expected that damping would be negligible for phase velocities near the velocity of light $c$. Therefore at the beginning we clearly delineate parameter regimes where either the phenomenon of Relativistic Landau Damping (regime 1) or the phenomenon of wave breaking (regime 2) would be dominant. In regime 1, we observe that the damping rate essentially follows the relativistic Landau damping rate derived first by Buti \cite{buti} in 1962. In the opposite regime (regime 2), we find that without any external perturbation the resultant wave propagates through the system for a long period of time and, in the low amplitude limit, it follows the relativistic warm plasma dispersion relation first given by Buti \cite{buti} and later derived by several other authors \cite{misra, pegoraro_porcelli, remi, melrose}. Further we find that, like a cold plasma Akhiezer - Polovin wave, in a warm plasma also relativistically intense wave breaks when perturbed by an arbitrarily small amplitude longitudinal perturbation. Breaking occurs at an amplitude far below the existing theoretical limits \cite{Katsouleas_88, rosenzweig_pra_88, scripta, meyer, schroeder_rapid, schroeder_06_POP, trines_2006} presented in the literature. We demonstrate that this breaking is a manifestation of the phase mixing phenomena \cite{sudip_prl, Sengupta_09, Sengupta_11, sudip_conference, prabal_main}, as mentioned above. We clearly show that after adding the external perturbation the characteristic frequency of the wave indeed becomes an explicit function of space which lead to wave breaking via phase mixing at an amplitude which is well below the existing theoretical limits. Further in section \ref{c5section3} we show that the results obtained from simulation indicate that the phase mixing time scale in a warm plasma can be interpreted using Dawson's formula \cite{dawson_main} for a non-relativistic cold inhomogeneous plasma, which is based on out of phase motion of neighbouring oscillators constituting the wave and separated by a distance equal to twice the amplitude of the oscillation/wave. Finally in section \ref{c5section4} we summarize this work and conclude.   

\section{Relativistic Particle-in-Cell Simulations Results} \label{c5section2}
In this section we perform PIC simulations with periodic boundary conditions in order to study the effect of finite electron temperature on the maximum electric field amplitude that can be sustained by a relativistically intense electron plasma wave in a warm plasma. For this purpose we first load Akhiezer - Polovin \cite{AP} type initial conditions in our relativistic PIC code. Along with this a finite temperature is also added to the electrons by loading a J\'{u}ttner - Synge \cite{synge} velocity distribution which can be expressed as (loaded using inversion method \cite{bird_book}):

\begin{equation}
f(p) = \frac{1}{2mcK_1(\frac{mc^2}{k_BT_e})} \exp \left[ -\frac{mc^2}{k_BT_e}\sqrt{1 + \frac{p^2}{m^2c^2}}\right]    \label{eq51}
\end{equation}
Here $K_1$ is the modified Bessel function of second kind \cite{abra}. In terms of $\lambda$, Eq.(\ref{eq51}) can be written as
\begin{equation}
f(p) = \frac{1}{2mcK_1(\frac{3}{\lambda})} \exp \left[ -\frac{3}{\lambda}\sqrt{1 + \frac{p^2}{m^2c^2}}\right]    \label{eq52}
\end{equation}

Ions are assumed to be infinitely massive providing a neutralizing positive background. Our simulation parameters are as follows: total number of particles $N_p = 80,000$, number of grid points $N_G = 500$, time step $\Delta t = \pi/160$. We use periodic boundary condition where the wavelength $L$ depends on the amplitude of the Akhiezer - Polovin wave which in turn is decided by its input parameters $u_m$ and $\beta _{\phi}$. Normalizations are as follows: $x \rightarrow x\omega _p/c$, $t \rightarrow \omega _pt$, $n_e \rightarrow n_e/n_0$, $v \rightarrow v/c$, $p \rightarrow p/mc$, $E \rightarrow eE/m\omega_p c$. In the following two subsections, we present the results obtained from the relativistic PIC simulations carried out in the respective regimes where the relativistic Landau damping (regime 1) $\&$ wave breaking (regime 2) are dominant.

\subsection{Regime 1 - Relativistic Landau Damping}
We have already mentioned that our goal is to verify the existing theoretical results \cite{Katsouleas_88, rosenzweig_pra_88, scripta, meyer, schroeder_rapid, schroeder_06_POP, trines_2006} on the maximum electric field amplitude that can be sustained by a relativistically intense wave in a warm plasma. For this purpose at the outset we should ensure ourself that the other effects would not interfere with the wave propagation. As the plasma under investigation contains a non-zero electron temperature, therefore it is possible that the temperature effect would try to damp the wave. As the waves are relativistically intense, here the reason for damping would be Relativistic Landau damping - first discovered by Buti \cite{buti} in 1962. By linearising relativistic Vlasov - Poisson's equations Buti \cite{buti} first wrote down the relativistic Landau damping rate $(\epsilon)$, which in the limit $k_BT_e \gg mc^2$ can be expressed as \cite{buti}
\begin{equation}
\epsilon =   -   \frac{1}{4}\pi ck \left[ 1 - \frac{3\lambda}{4} \right]    \label{eqff}
\end{equation}
 The author \cite{buti} also mentioned that, ``the damping is very strong in the case where phase velocity is small compared to $c$''.  
\begin{figure}[htbp]
\begin{center}
\includegraphics[scale=0.4]{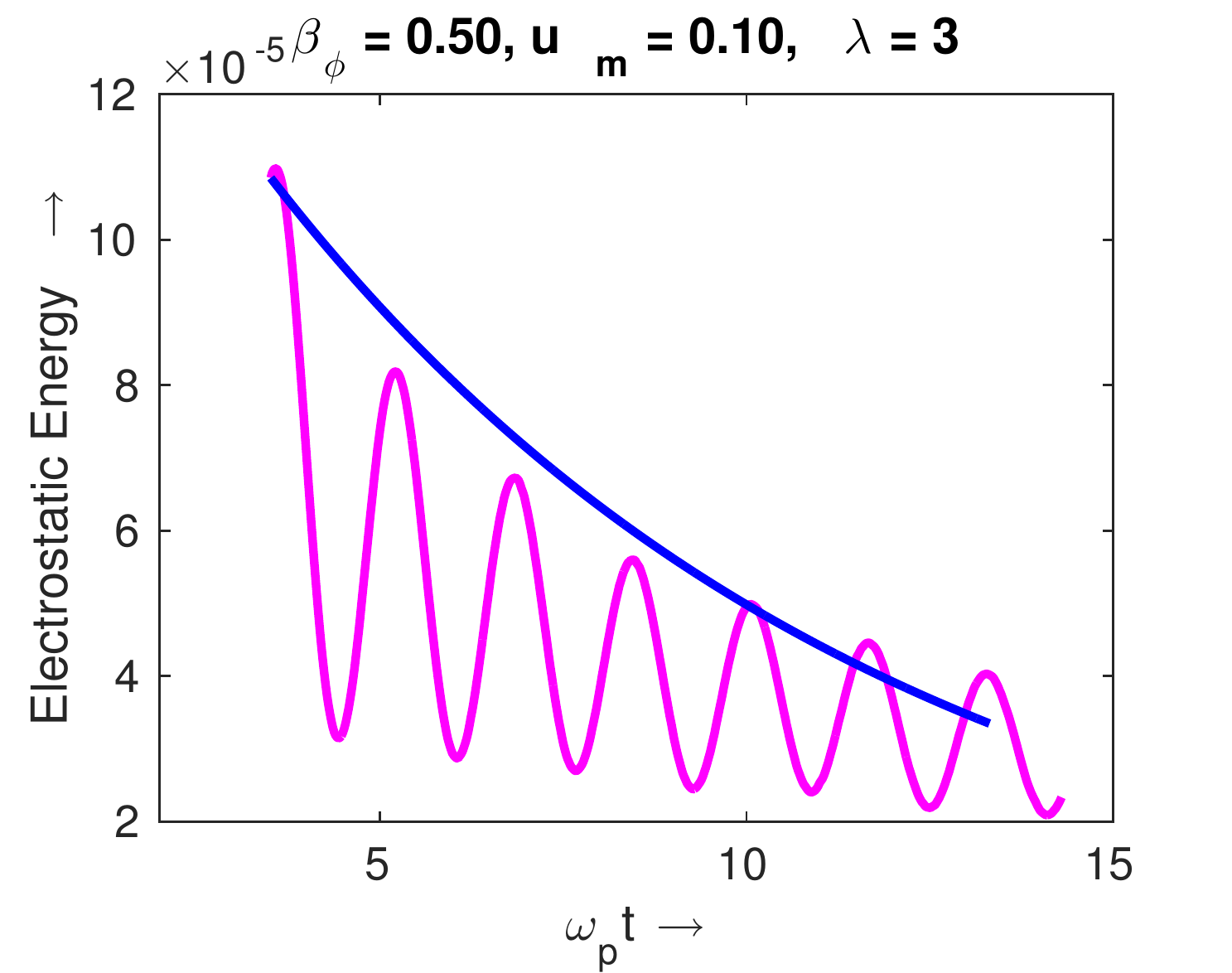}
\caption{Time evolution of electrostatic energy at a fixed grid point for $\beta _{\phi} = 0.5$, $u_m = 0.1$, $\lambda = 3$}
\label{fig5c}
\end{center}
\end{figure}
In Fig.(\ref{fig5c}), we have shown the time evolution of electrostatic energy obtained from simulation for $\beta _{\phi} = 0.5$, $u_m = 0.1$, $\lambda = 3$ in magenta colour. The blue line is the relativistic landau damping rate derived by Buti \cite{buti}. This figure shows that the damping rate follows the theoretical predictions made by Buti \cite{buti} which is given by Eq.(\ref{eqff}). This damping rate decreases significantly as $\beta _{\phi}$ increases. From the complete analysis of relativistic Landau damping rate it is shown that the damping rate is very small for wave Lorentz factor $\gamma _\phi \gg 1 + \lambda/3$ \cite{buti, melrose}. When this inequality is reversed the damping becomes very strong and swamps out the wave propagation and hence the wave breaking physics. 

\begin{figure}[htbp]
\begin{center}
\includegraphics[scale=0.45]{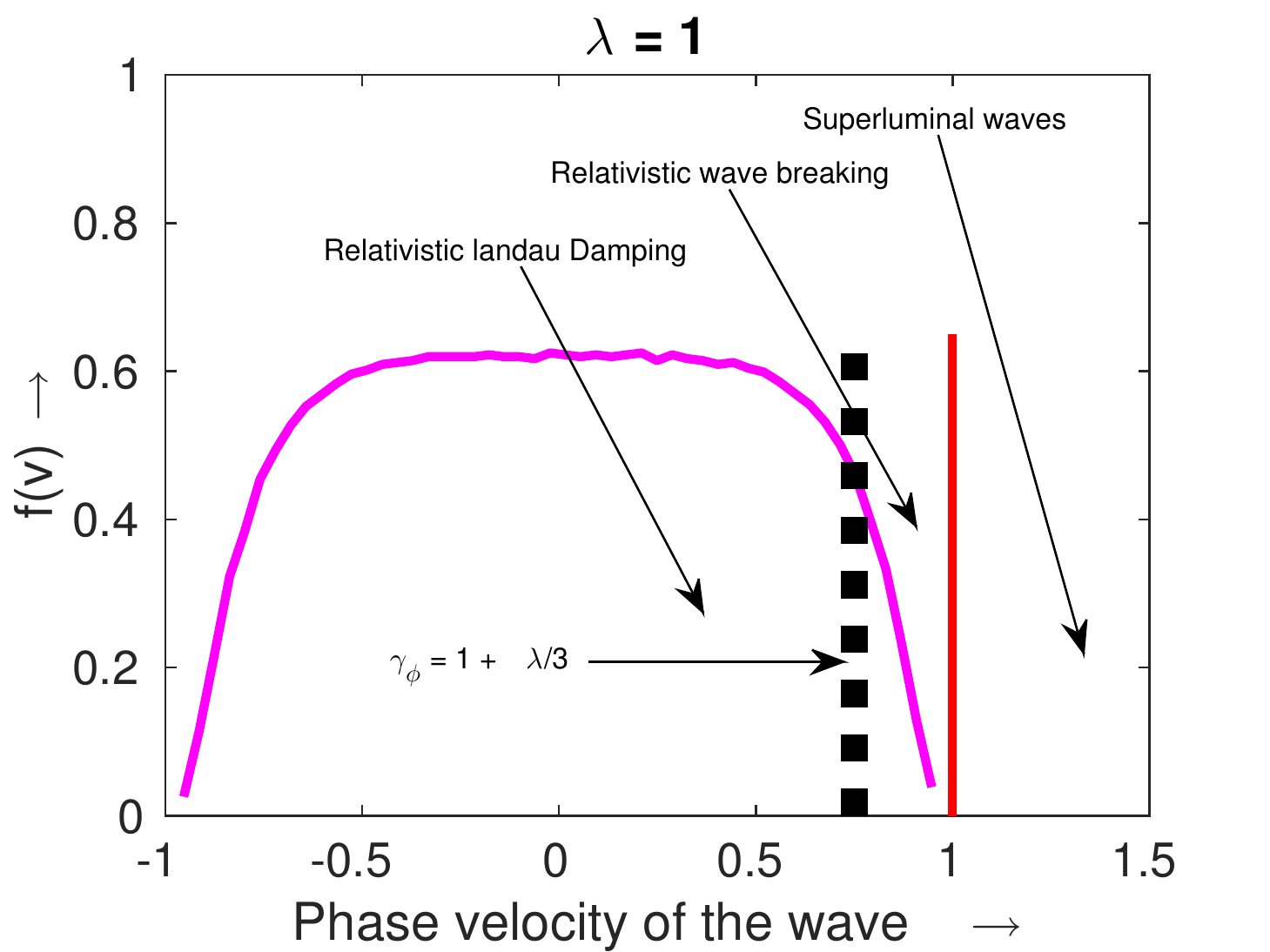}
\caption{Parameter domains exhibiting relativistic Landau damping and wave breaking}
\label{fig5d}
\end{center}
\end{figure}

For the sake of clarity, here we have shown a schematic diagram of relativistic velocity distribution function [Fig.(\ref{fig5d})] where we have roughly depicted the parameter regimes (for $\lambda = 1$) where either relativistic Landau damping or wave breaking would be dominant. From this figure we understand that to study wave breaking we need to work in the regime between the vertical black (corresponding to $\beta _\phi = 0.68$, for $\gamma _\phi = 1 + \lambda/3$) and the red line (corresponding to $\beta _\phi = 1$) where wave particle interaction is almost negligible. Therefore in the next subsection to explore the wave breaking physics, we keep the phase velocity of the Akhiezer - Polovin waves $\beta _{\phi} = 0.95$ such that the relativistic Landau Damping rate remains small for the entire range of $\lambda$ where we carry out the numerical experiment.

\subsection{Regime 2 - Wave breaking (via phase mixing)}
\begin{figure}[htbp]
\begin{center}
\includegraphics[scale=0.35]{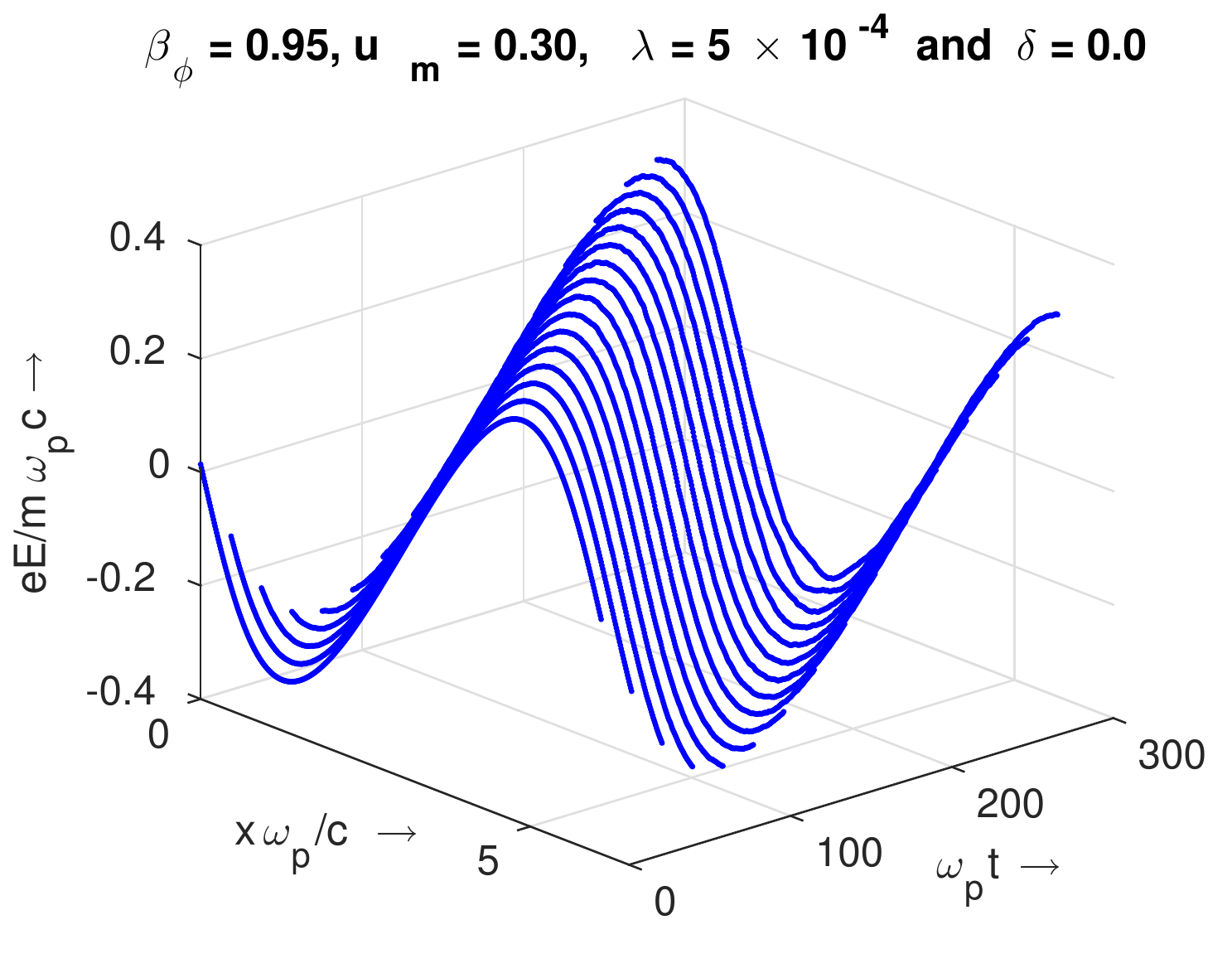}
\caption{Space-time evolution of electric field for $\beta _{\phi} = 0.95$, $u_m = 0.3$, $\lambda = 5\times 10^{-4}$ and $\delta = 0.0$}
\label{fig5e1}
\end{center}
\end{figure}

\begin{figure}[htbp]
\begin{center}
\includegraphics[scale=0.35]{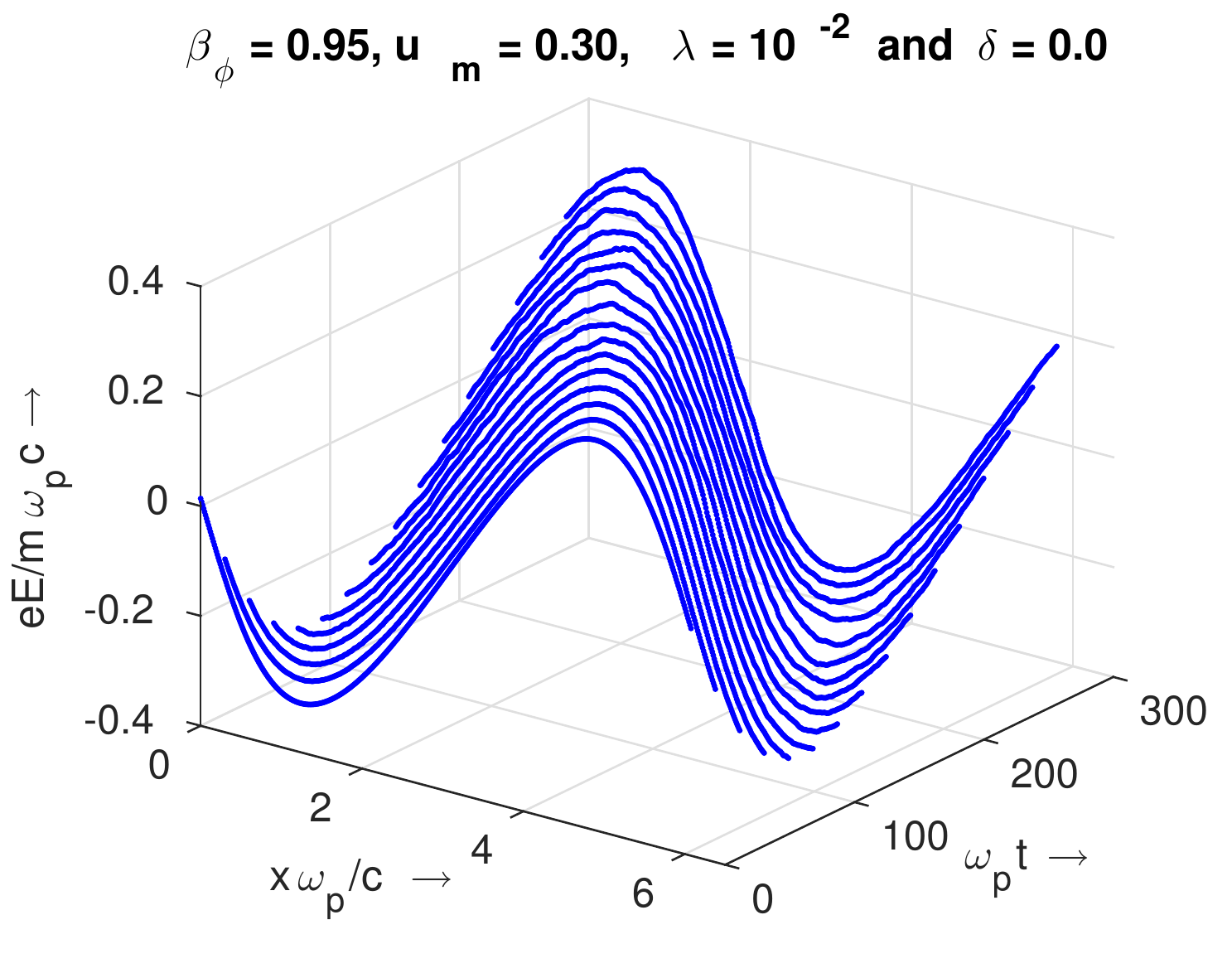}
\caption{Space-time evolution of electric field for $\beta _{\phi} = 0.95$, $u_m = 0.3$, $\lambda = 10^{-2}$ and $\delta = 0.0$}
\label{fig5e2}
\end{center}
\end{figure}

\begin{figure}[htbp]
\begin{center}
\includegraphics[scale=0.25]{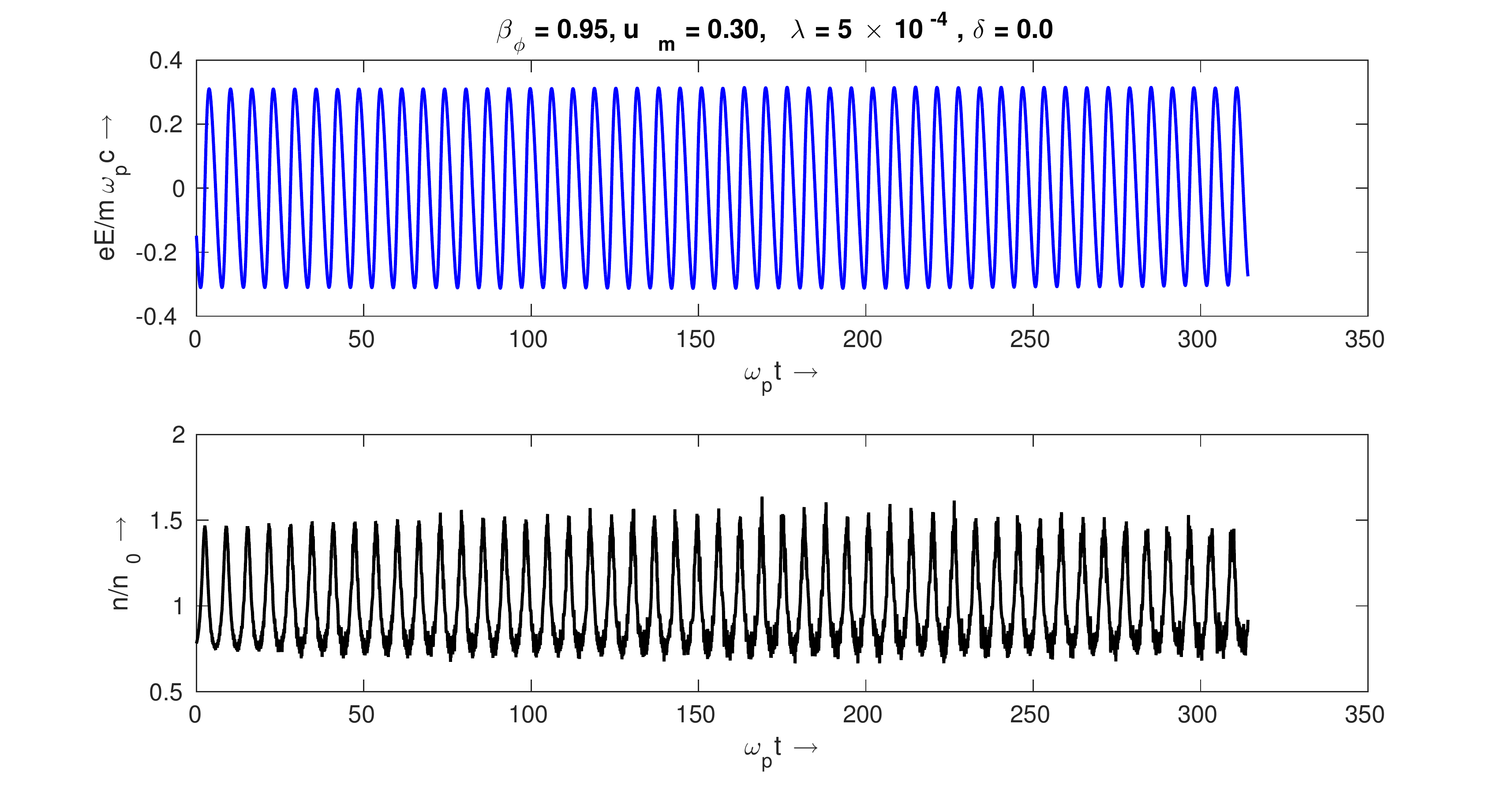}
\caption{Time evolution of electric field $\&$ density at a fixed grid point for $\beta _{\phi} = 0.95$, $u_m = 0.3$, $\lambda = 5\times 10^{-4}$ and $\delta = 0.0$}
\label{fig5f1}
\end{center}
\end{figure}

\begin{figure}[htbp]
\begin{center}
\includegraphics[scale=0.25]{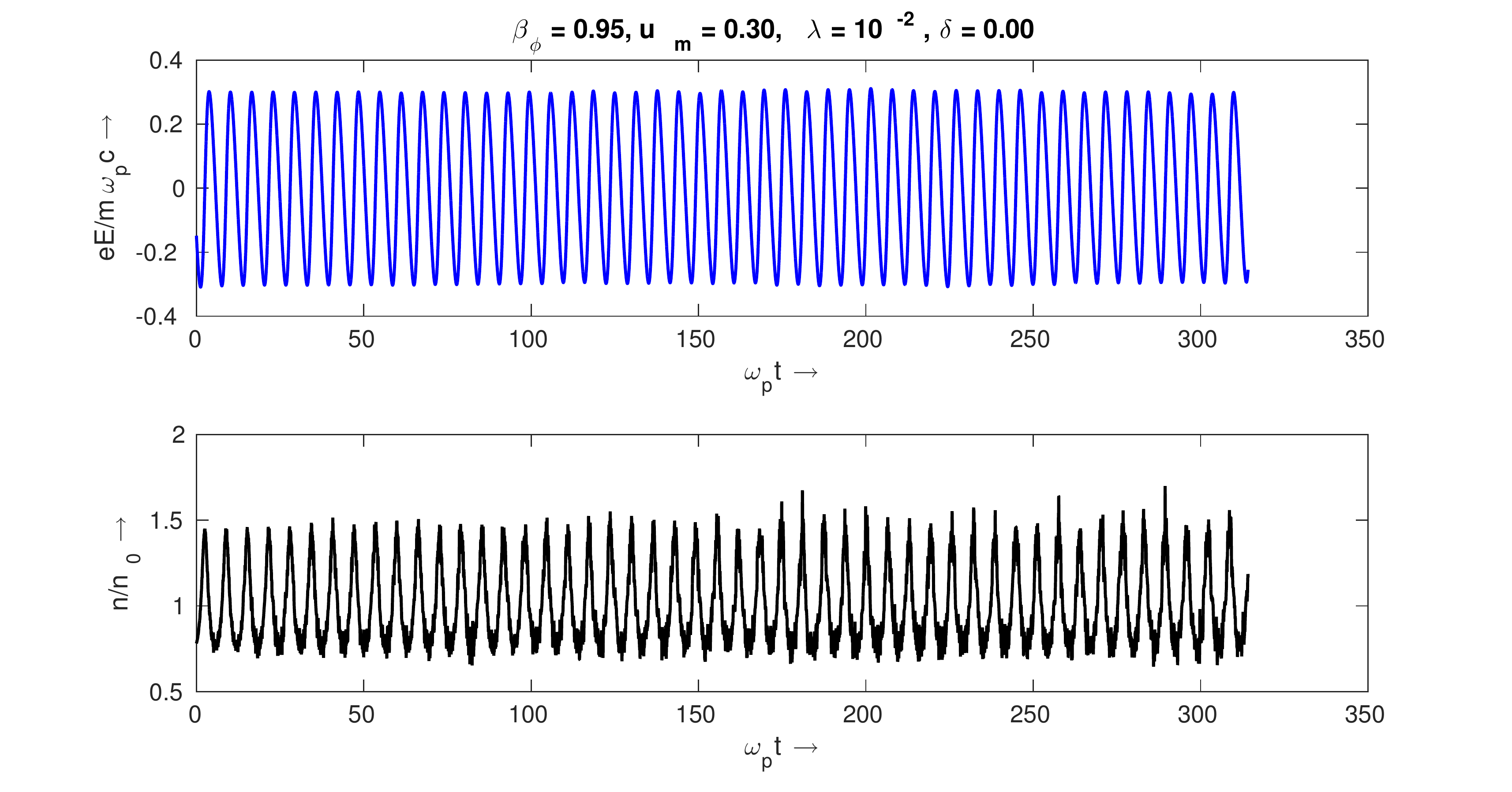}
\caption{Time evolution of electric field $\&$ density at a fixed grid point for $\beta _{\phi} = 0.95$, $u_m = 0.3$, $\lambda = 0.01$ and $\delta = 0.0$}
\label{fig5f2}
\end{center}
\end{figure}

\begin{figure}[htbp]
\begin{center}
\includegraphics[scale=0.35]{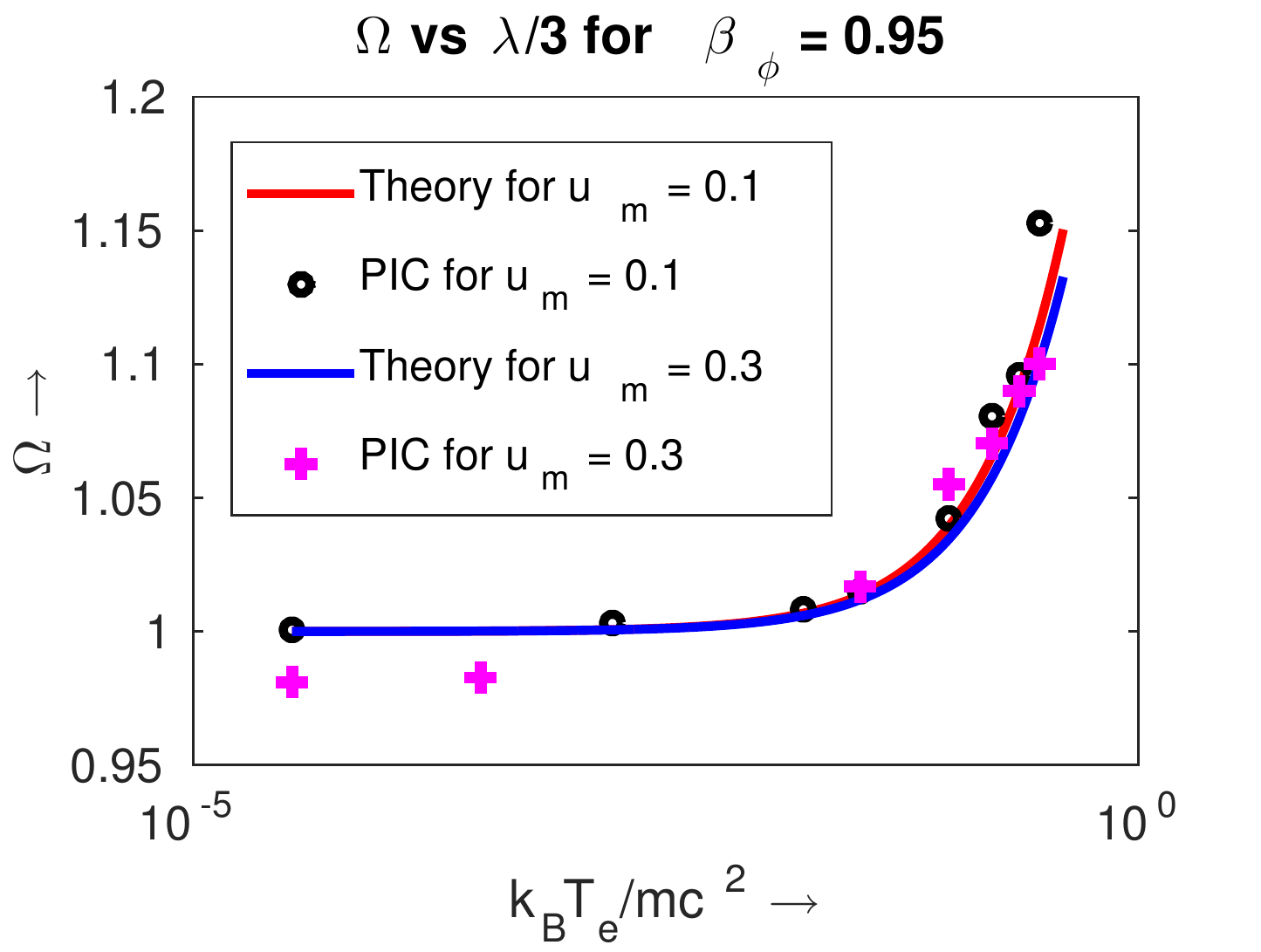}
\caption{$\Omega$ as a function of $\lambda$, for $\beta _{\phi} = 0.95$, $u_m = 0.10,0.30$, $\delta = 0.0$}
\label{fig5g}
\end{center}
\end{figure}

\begin{figure}[htbp]
\begin{center}
\includegraphics[scale=0.35]{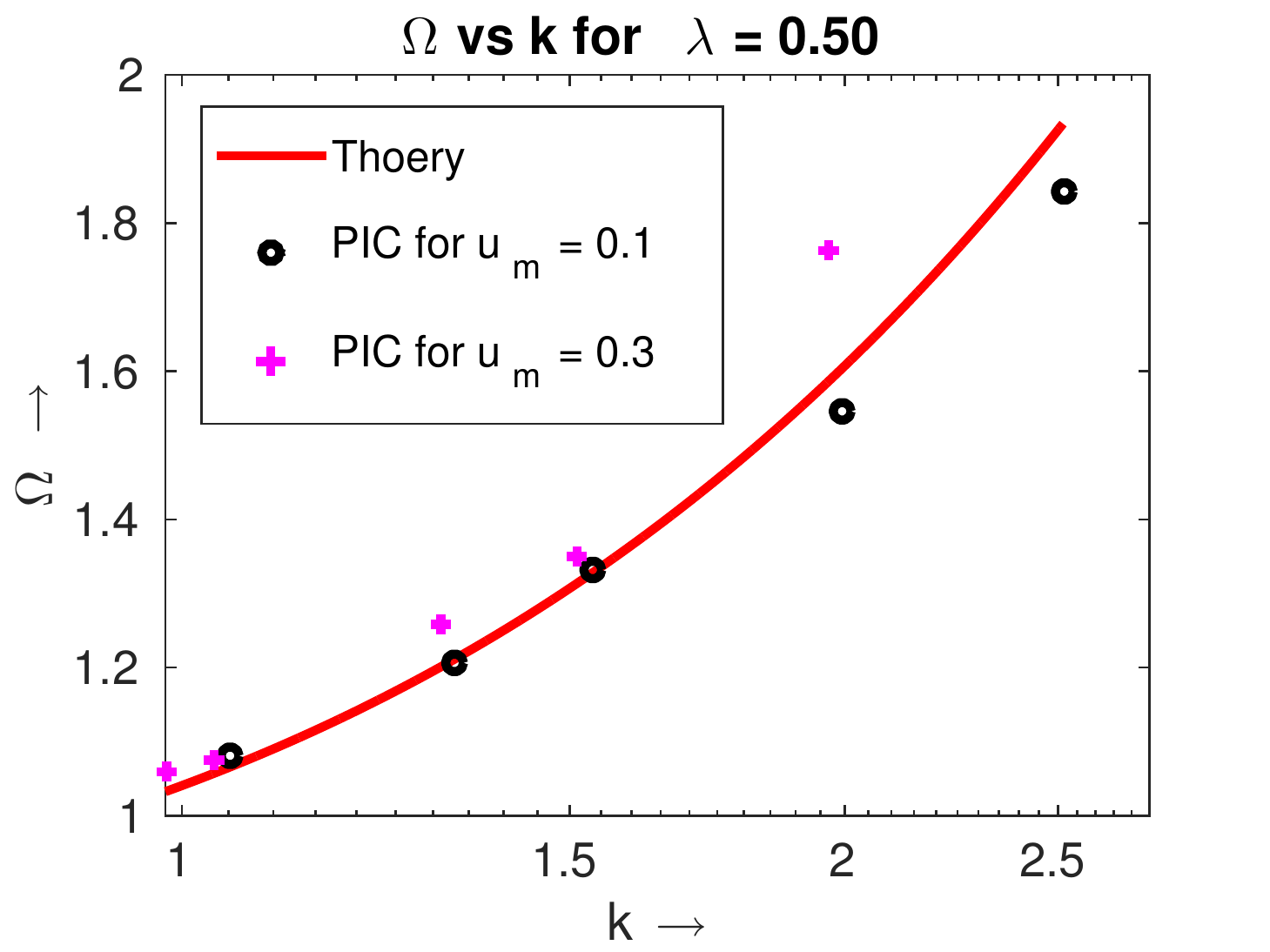}
\caption{$\Omega$ as a function of $k$, for $u_m = 0.10, 0.30$, $\delta = 0.0$ ($\lambda$ is taken as $0.5$)}
\label{fig5h}
\end{center}
\end{figure}

 Here, in all simulation runs we keep the phase velocity of the Akhiezer - Polovin wave at $\beta _{\phi} = 0.95$. Thus the relativistic Landau damping rate is negligible and the wave should propagate for a long period of time without damping or loosing the periodicity. Figs.(\ref{fig5e1}) and (\ref{fig5e2}) respectively show the space time evolution of the electric field profile of a relativistically intense wave for $\lambda = 5\times 10^{-4}$ and $\lambda = 10^{-2}$. The value of $u_m$ is taken as $0.30$. From these figures we see that the wave propagates through the system without any damping and without loosing periodicity for a large period of time. In Figs.(\ref{fig5f1}) and (\ref{fig5f2}), we have also plotted the time evolution of the electric field $\&$ density at a fixed grid point for two different initial temperature and observe that both are oscillating with a nearly constant amplitude. By measuring the time difference between two consecutive peaks from Figs.(\ref{fig5f1}) and (\ref{fig5f2}) we find that when a finite temperature is added with the pure Akhiezer - Polovin wave, the resultant frequency ($\Omega$) does not remain as the frequency of cold plasma Akhiezer - Polovin wave ($\Omega _{ap}$)\cite{AP, prabal_main, sudip_conference}. In the small amplitude limit, we can estimate this characteristic frequency from the relativistic warm plasma dispersion relation \cite{buti, misra, remi, melrose} which is followed by a relativistically intense electron plasma wave in a warm plasma. In the limit $k_BT_e/mc^2 << 1$ this dispersion relation can be written as 
\begin{equation}
\Omega ^2 = \omega _p^2 + k^2c^2\lambda - \frac{5}{6}\omega _p^2\lambda  \label{eq53}
\end{equation}
 Here, we verify this dispersion relation for two different amplitudes $u_m = 0.1$ $\&$ $0.3$ by changing the values of $\lambda$ and $k$. In Figs.(\ref{fig5g}) and (\ref{fig5h}), we respectively show the variation of frequency $\Omega$ as a function of $\lambda$ and $k$ for fixed value of other parameters. In these figures the points are obtained from PIC simulations and the continuous lines are the theoretical relativistic dispersion relation given by Eq.(\ref{eq53}). Note that, for amplitude $u_m= 0.1$ we see a better matching as compared to $u_m = 0.3$. This is expected, as this dispersion relation is obtained by linearising the Valsov - Poisson's equations, the excited wave is supposed to follow it only in the low amplitude limit. 


\begin{figure}[htbp]
\begin{center}
\includegraphics[scale=0.35]{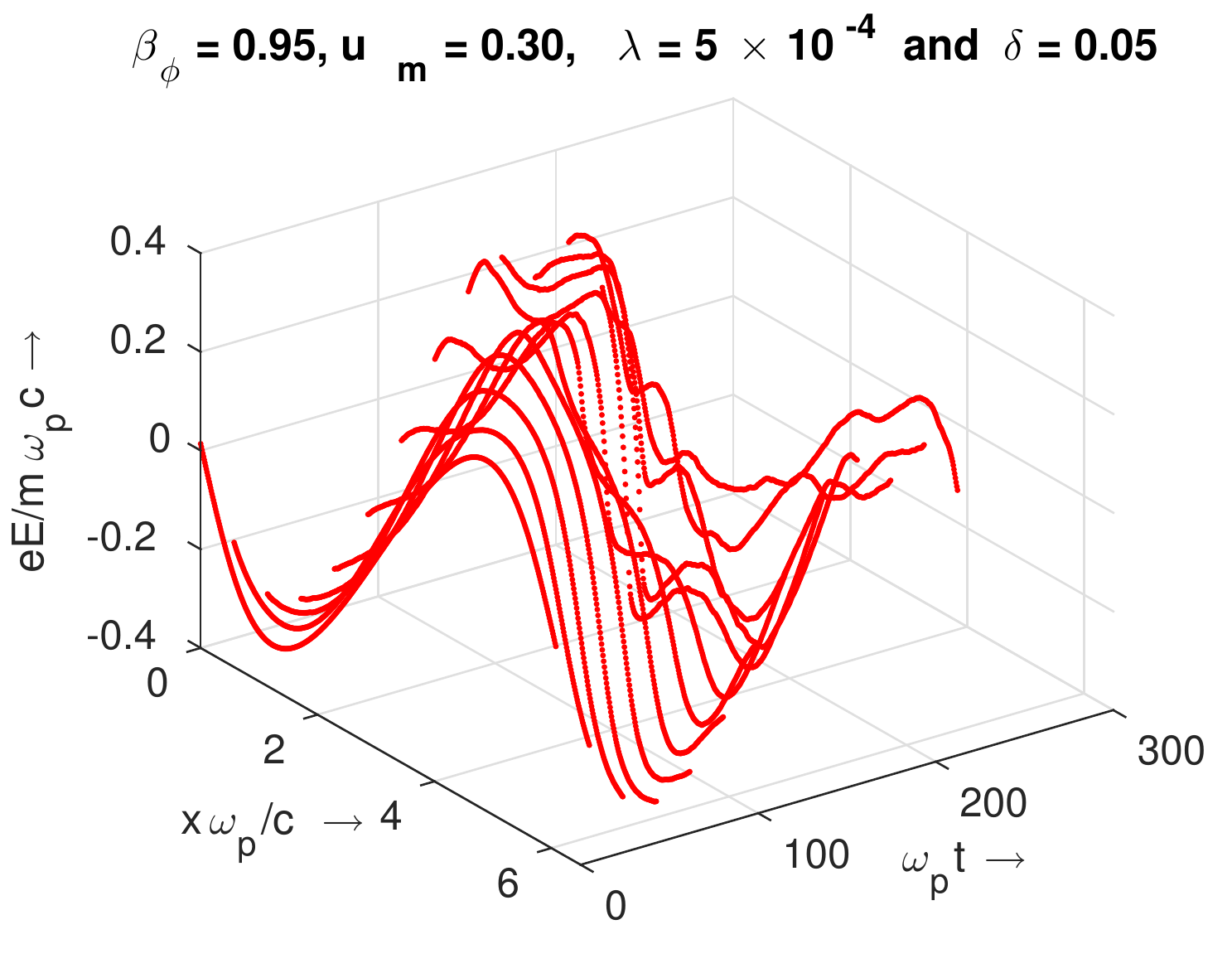}
\caption{Space-time evolution of electric field for $\beta _{\phi} = 0.95$, $u_m = 0.3$, $\lambda = 5\times 10^{-4}$ and $\delta = 0.05$}
\label{fig5i1}
\end{center}
\end{figure}
\begin{figure}[htbp]
\begin{center}
\includegraphics[scale=0.35]{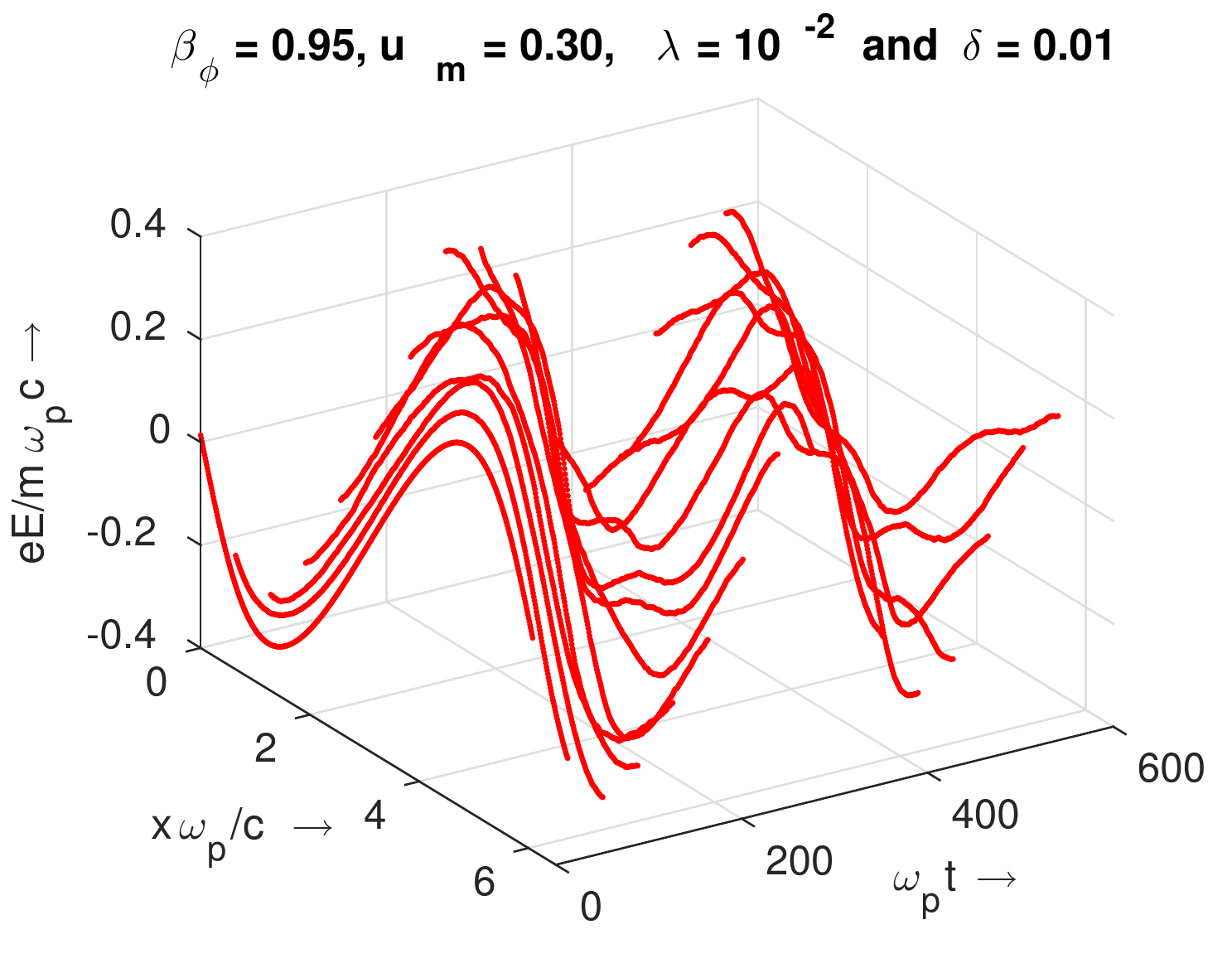}
\caption{Space-time evolution of electric field for $\beta _{\phi} = 0.95$, $u_m = 0.3$, $\lambda = 10^{-2}$ and $\delta = 0.01$}
\label{fig5i2}
\end{center}
\end{figure}

Now we add a very small amplitude sinusoidal velocity perturbation with a maximum amplitude $\delta$ to this large amplitude Akhiezer - Polovin wave with same mode number as the large amplitude Akhiezer - Polovin wave ($k_{ap}$). In Figs.(\ref{fig5i1}) and (\ref{fig5i2}) we show that the space -time evolution of the resultant electric field of the perturbed wave for two different values of $\lambda$ and $\delta$. 
\begin{figure}[htbp]
\begin{center}
\includegraphics[scale=0.25]{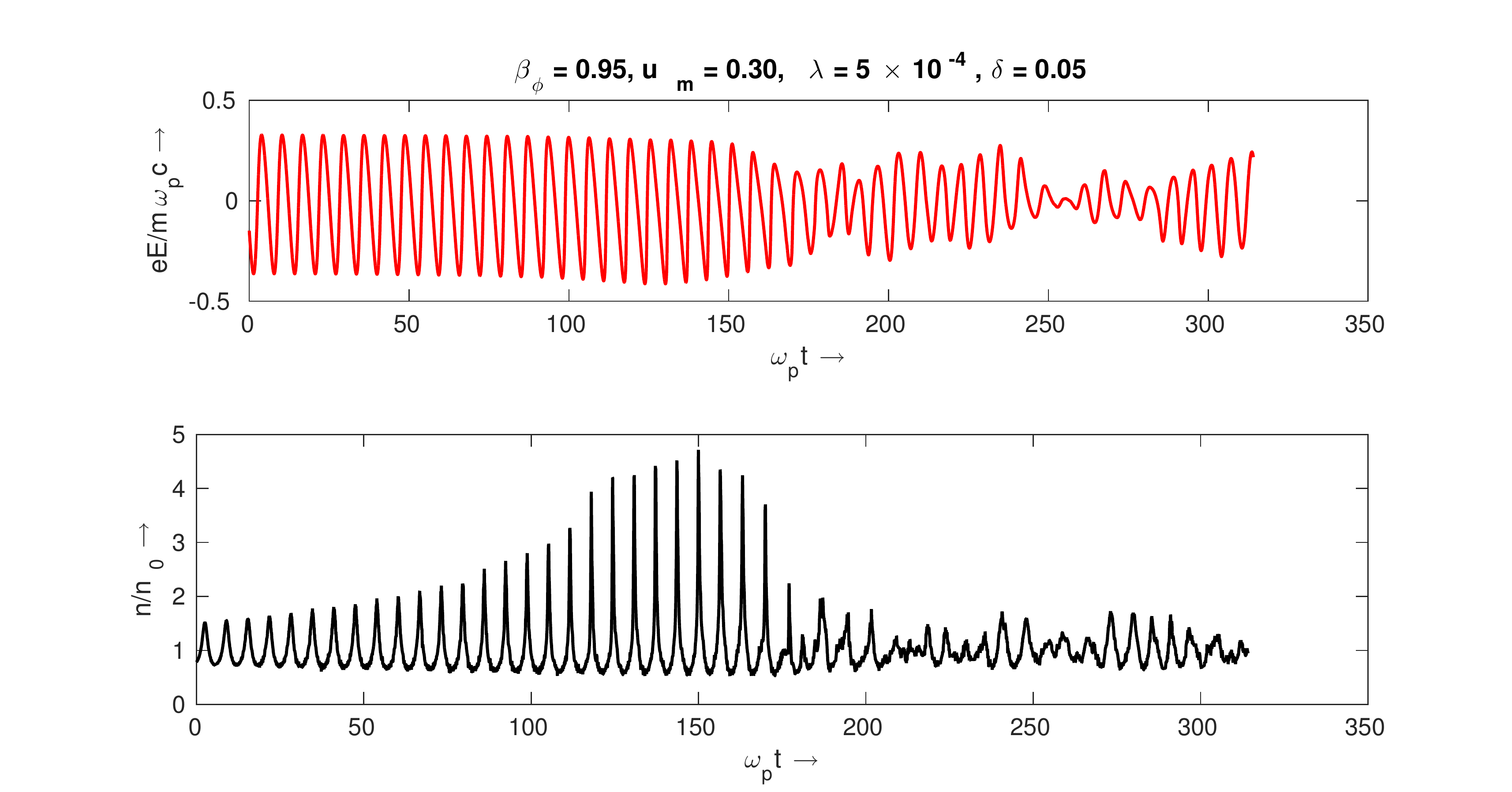}
\caption{Time evolution of electric field $\&$ density at a fixed grid point for $\beta _{\phi} = 0.95$, $u_m = 0.3$, $\lambda = 5\times 10^{-4}$ and $\delta = 0.05$}
\label{fig5m1}
\end{center}
\end{figure}
\begin{figure}[htbp]
\begin{center}
\includegraphics[scale=0.25]{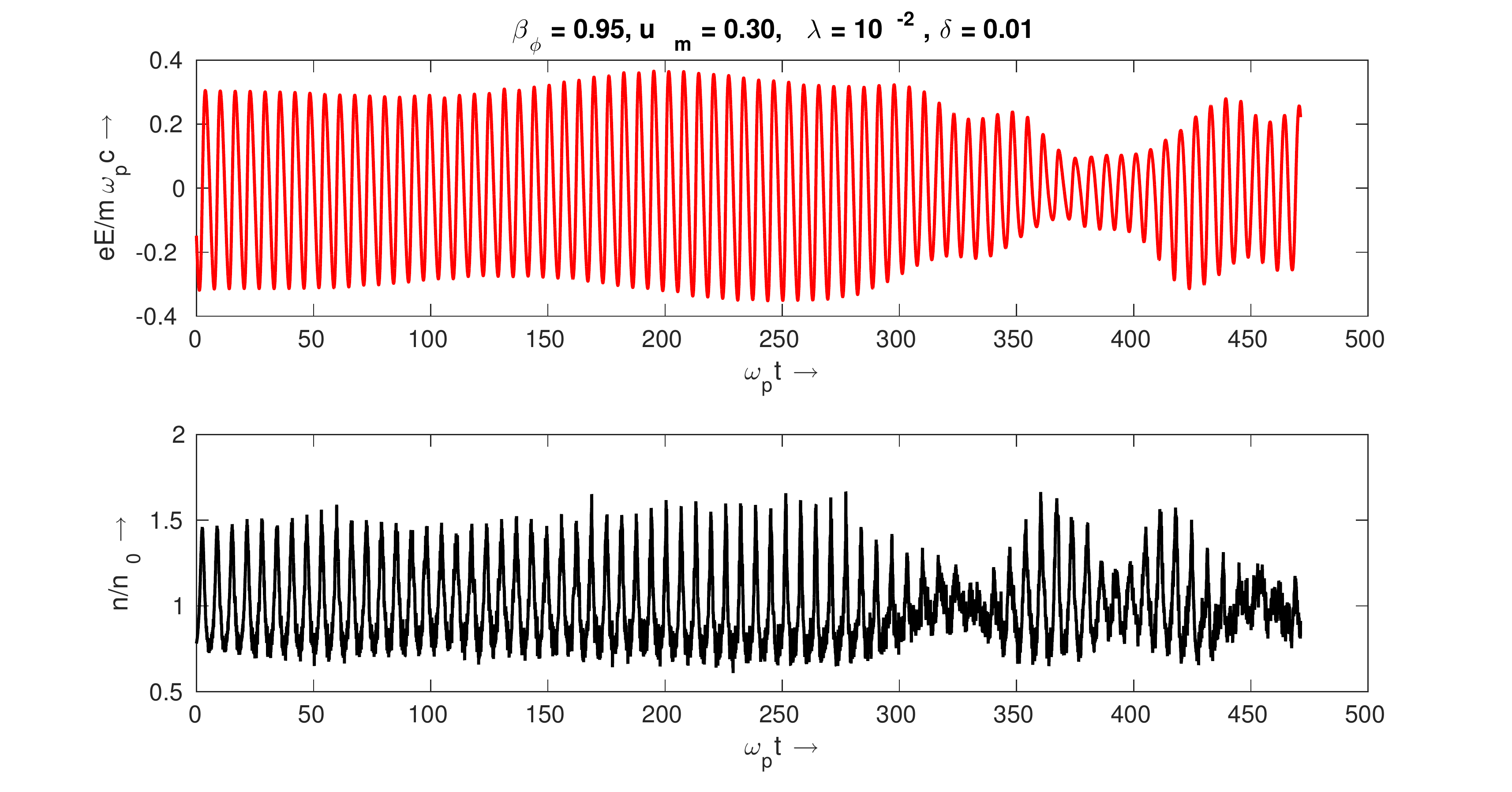}
\caption{Time evolution of electric field $\&$ density at a fixed grid point for $\beta _{\phi} = 0.95$, $u_m = 0.3$, $\lambda = 10^{-2}$ and $\delta = 0.01$}
\label{fig5m2}
\end{center}
\end{figure} 
Similarly, in Figs.(\ref{fig5m1}) and (\ref{fig5m2}), we have plotted the time evolution of the perturbed electric field and density at a fixed grid point for the same parameters. Both the figures are exhibiting gradual deformation of the wave electric field and density profile which show that as time progresses the wave profile deforms and after a certain time (decided by $u_m$, $\beta _{\phi}$, $\lambda$ and $\delta$) the wave amplitude becomes modulated. We define wave breaking time (phase mixing time) as the time when the ``first dip'' appears in the time evolution plot [Figs.(\ref{fig5m1}) and \ref{fig5m2})]. We expect that this breaking is manifested via the process of phase mixing, as after adding the perturbation the characteristic frequency could become a function of space. To confirm our prediction we have measured the initial total energy ($``a"$) of each particle (electron sheet) for both the cases without and with perturbation. Then the characteristic frequencies of the motion of the particles have been evaluated by using the general expression of frequency for a relativistic harmonic oscillator, which is given by 
\begin{equation}
\Omega = \omega _{p}\frac{\pi}{2}\frac{r^ \prime}{[2E(r) - r^{\prime}K(r)]}   \label{eq54} 
\end{equation}
 where $E(r)$ and $K(r)$ are complete elliptic integrals of second and first kind \cite{abra} respectively, $r^2 = (a-1)/(a+1)$ and $r^ \prime = (1 - r^2)^{1/2}$. In Figs.(\ref{fig5x}) and (\ref{fig5y}) we have respectively plotted the frequency (averaged over a cell) as a function of space for the cases without and with perturbation for a fixed electron temperature ($\lambda = 0.01$). We observe that in Fig.(\ref{fig5y}), after adding the perturbation, the frequency indeed becomes a function of position as the total energy $``a"$ becomes an explicit function of the space; this is absent in Fig.(\ref{fig5x}) as for $\delta = 0$ the energy of each particle remains independent of their respective equilibrium position. 
\begin{figure}[htbp]
\begin{center}
\includegraphics[scale=0.32]{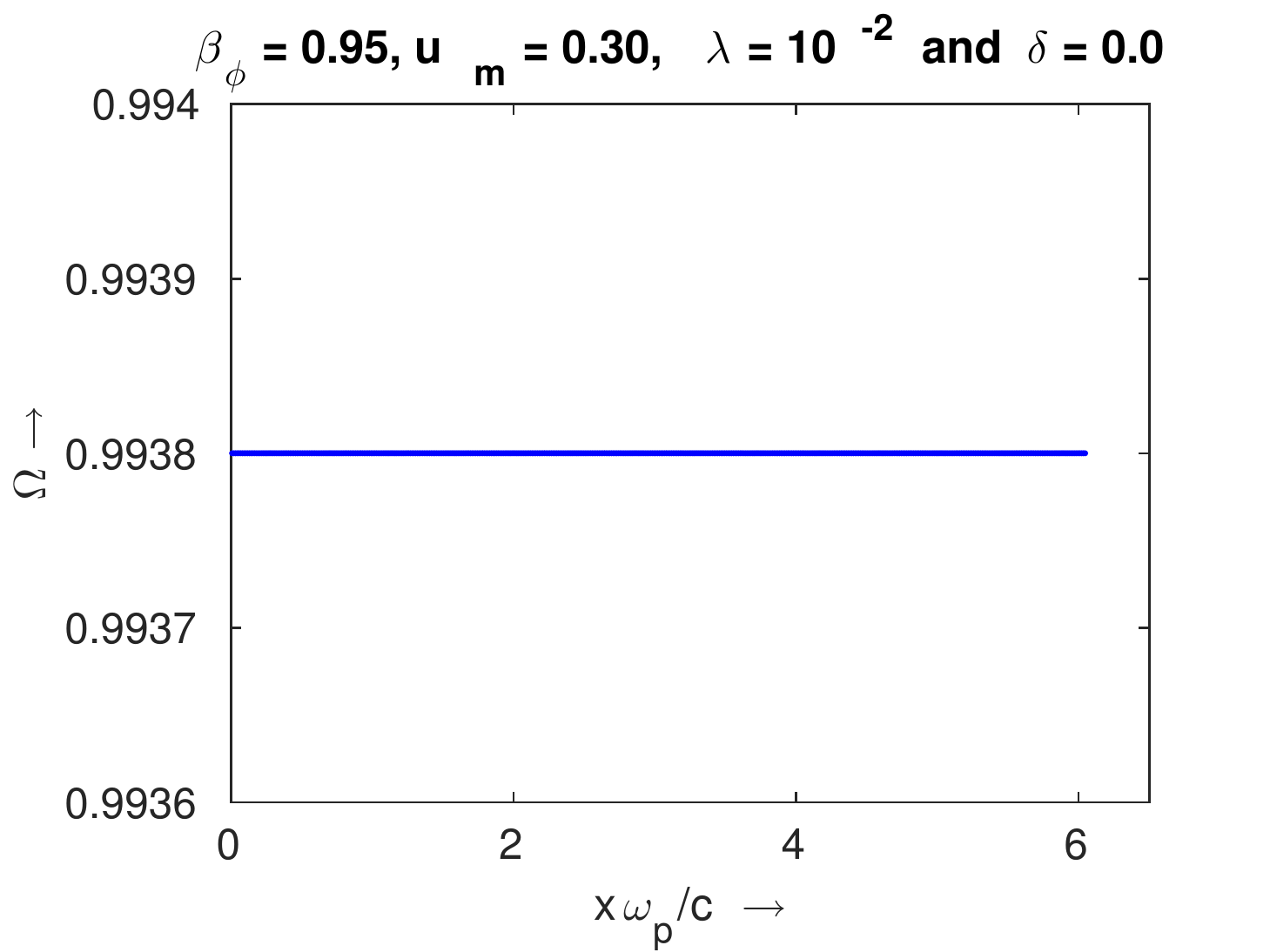}
\caption{Frequency of the wave as a function of the position for $\beta _{\phi} = 0.95$, $u_m = 0.3$, $\lambda = 0.01$ and $\delta = 0.0$}
\label{fig5x}
\end{center}
\end{figure} 
 
\begin{figure}[htbp]
\centering
\includegraphics[scale=0.29]{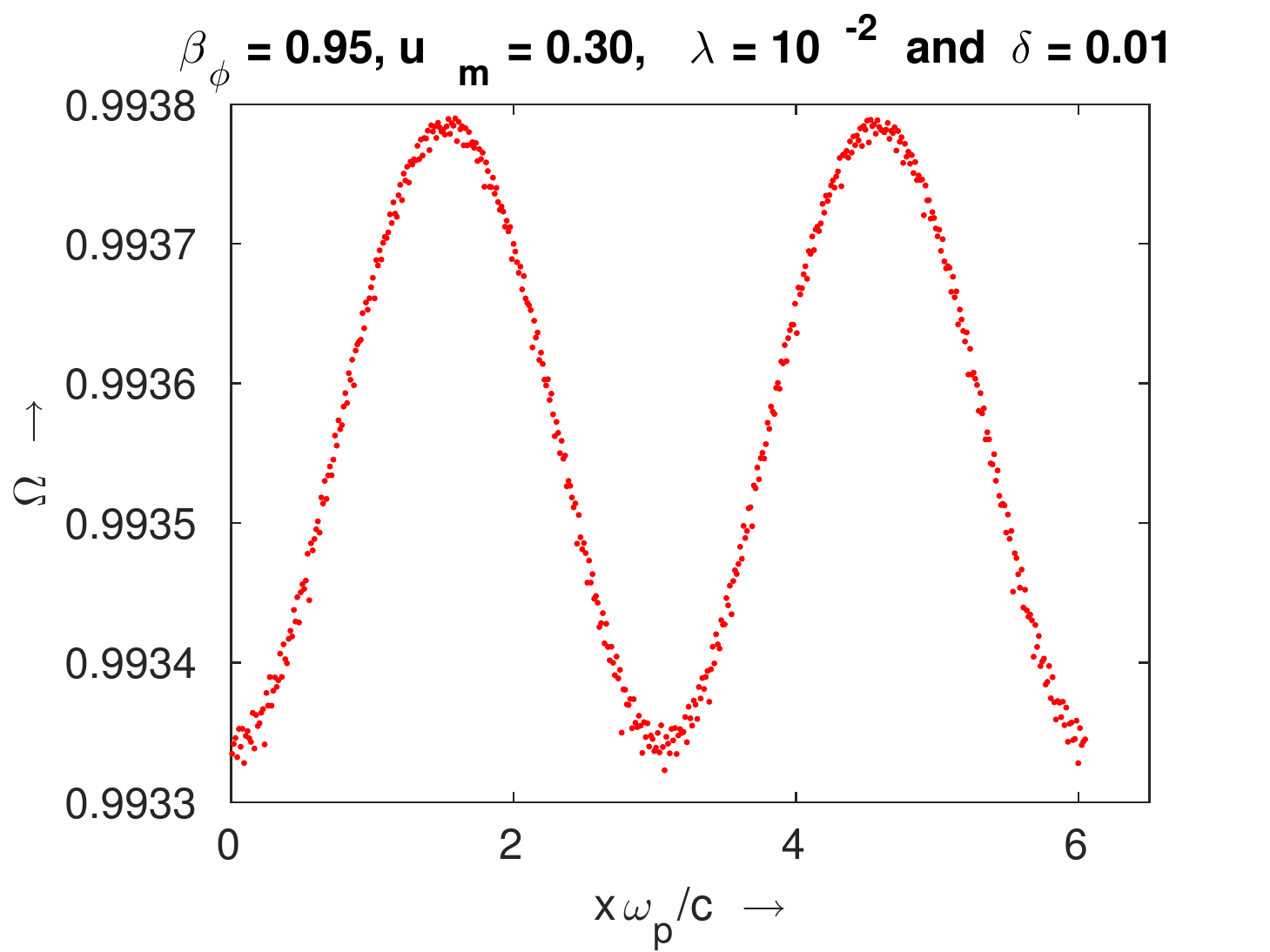}
\caption{Frequency of the wave as a function of the position for $\beta _{\phi} = 0.95$, $u_m = 0.3$, $\lambda = 0.01$ and $\delta = 0.01$}
\label{fig5y}
\end{figure}
\begin{figure}[htbp]
\centering
\includegraphics[scale=0.29]{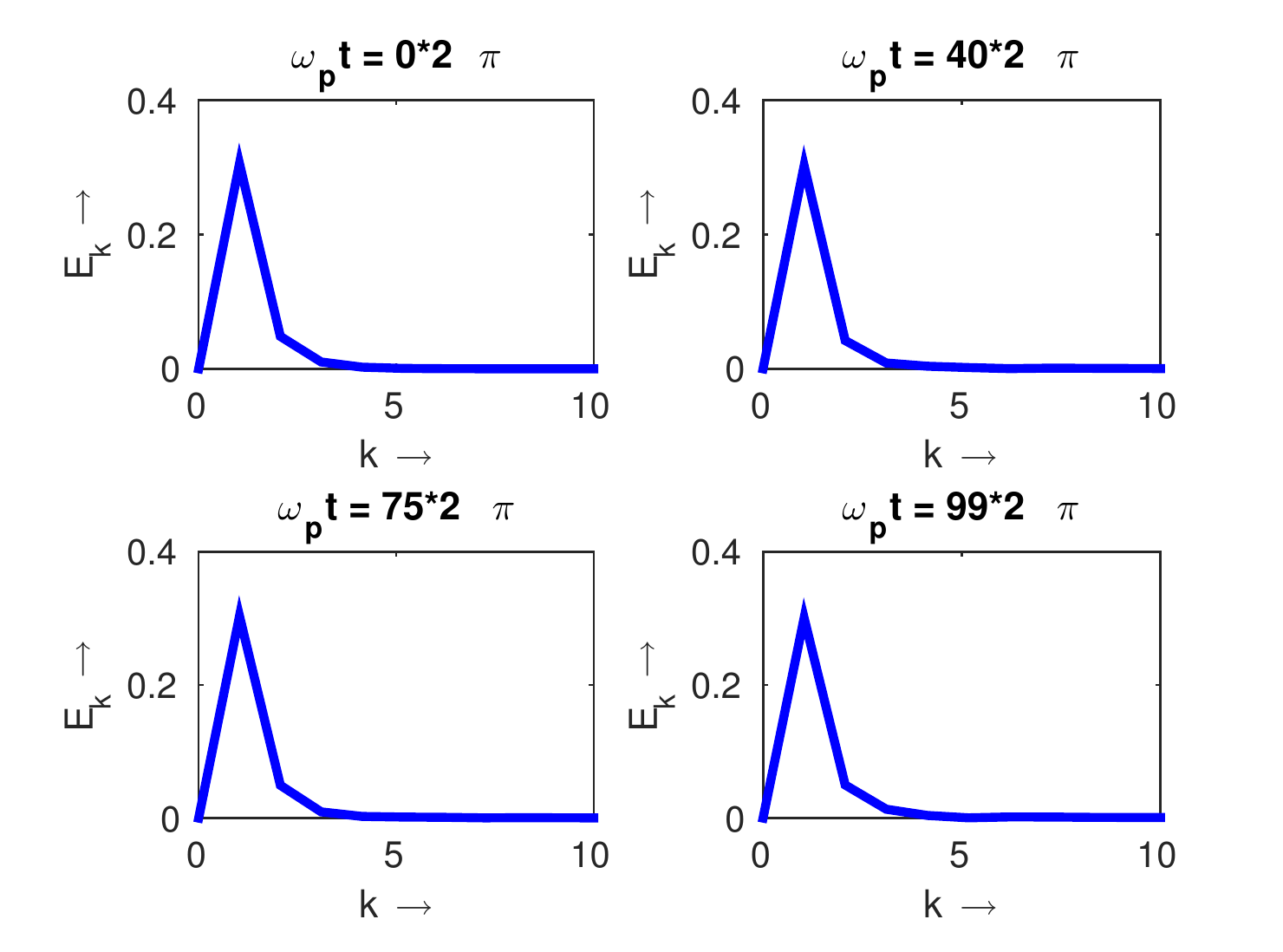}
\includegraphics[scale=0.28]{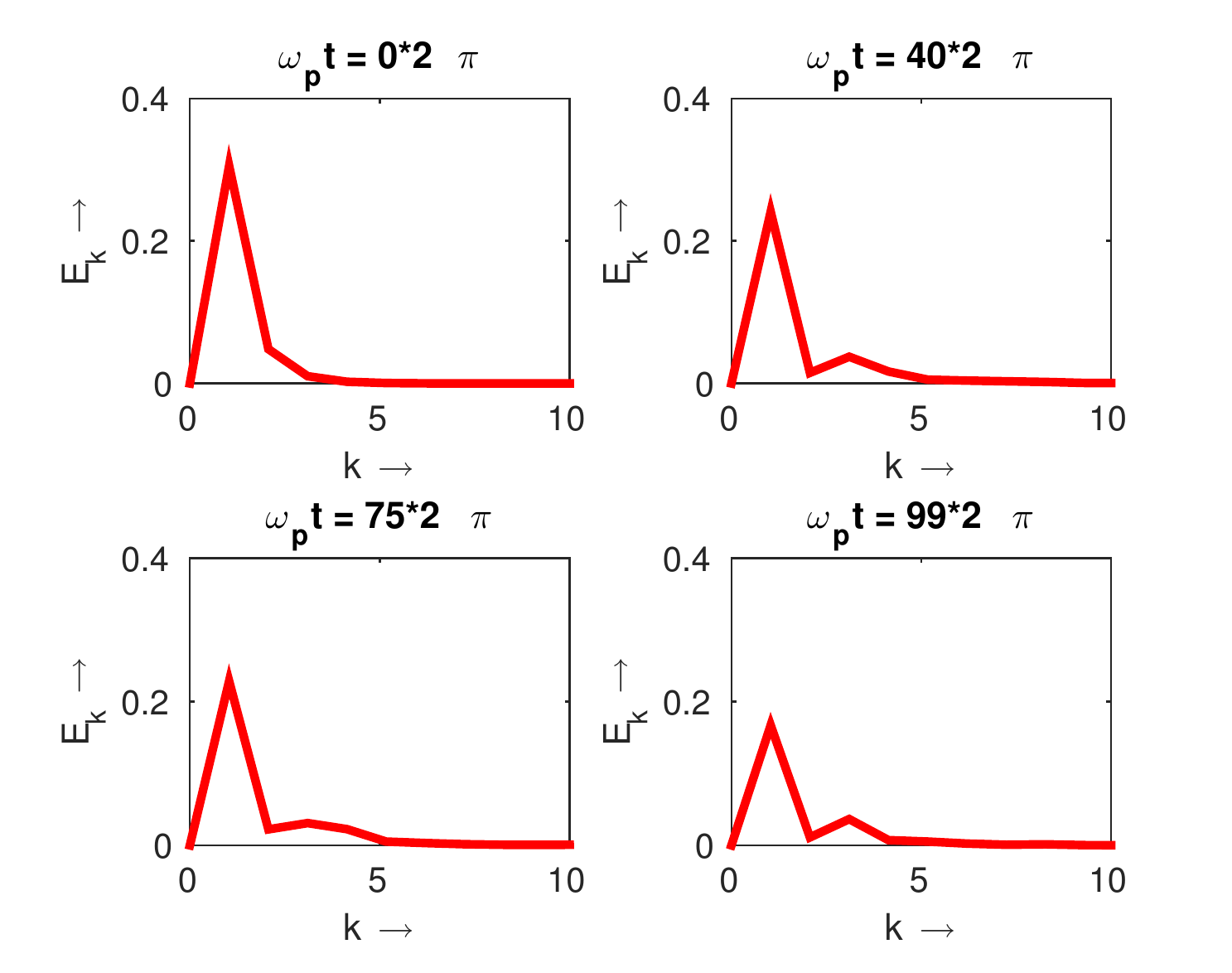}
\caption{Fourier spectrum of electric field for $\beta _{\phi} = 0.95$, $u_m = 0.3$, $\lambda = 0.01$ and $\delta = 0.0$ (blue), $0.01$(red) at different time steps}
\label{fig5l}
\end{figure}

The occurance of phase mixing has been again confirmed by plotting the Fourier spectrum of electric field amplitude ($E_k$) at different instants of time for both the cases without (left) and with perturbation (right). Compare these two figures (blue and red) in Fig.(\ref{fig5l}). It shows that, after adding the perturbation, as the time progresses the amplitude of the primary mode ($k_{ap}$) reduces significantly (red curves) with the simultaneous growth in higher order modes. It is clear from the Fig.(\ref{fig5l}) that a significant amount of energy has been transferred to the higher harmonics which is another signature of wave breaking via phase mixing process as reported by other authors \cite{sudip_prl, Sengupta_09, Sengupta_11}. As a consequence some of the electrons will acquire energy from the wave and accelerate to much higher energies. 
 
\begin{figure}[htbp]
\centering
\includegraphics[scale=0.24]{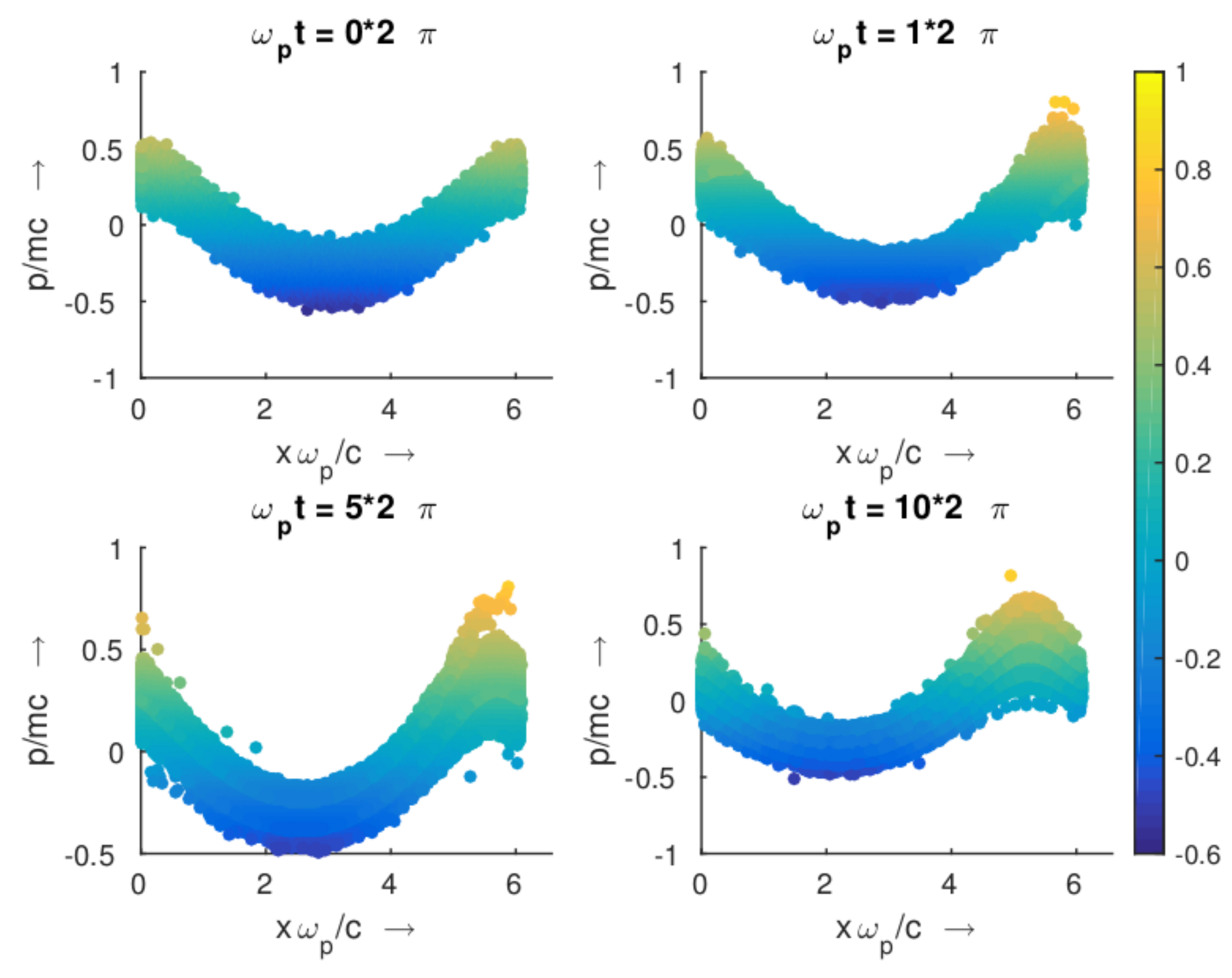}
\includegraphics[scale=0.24]{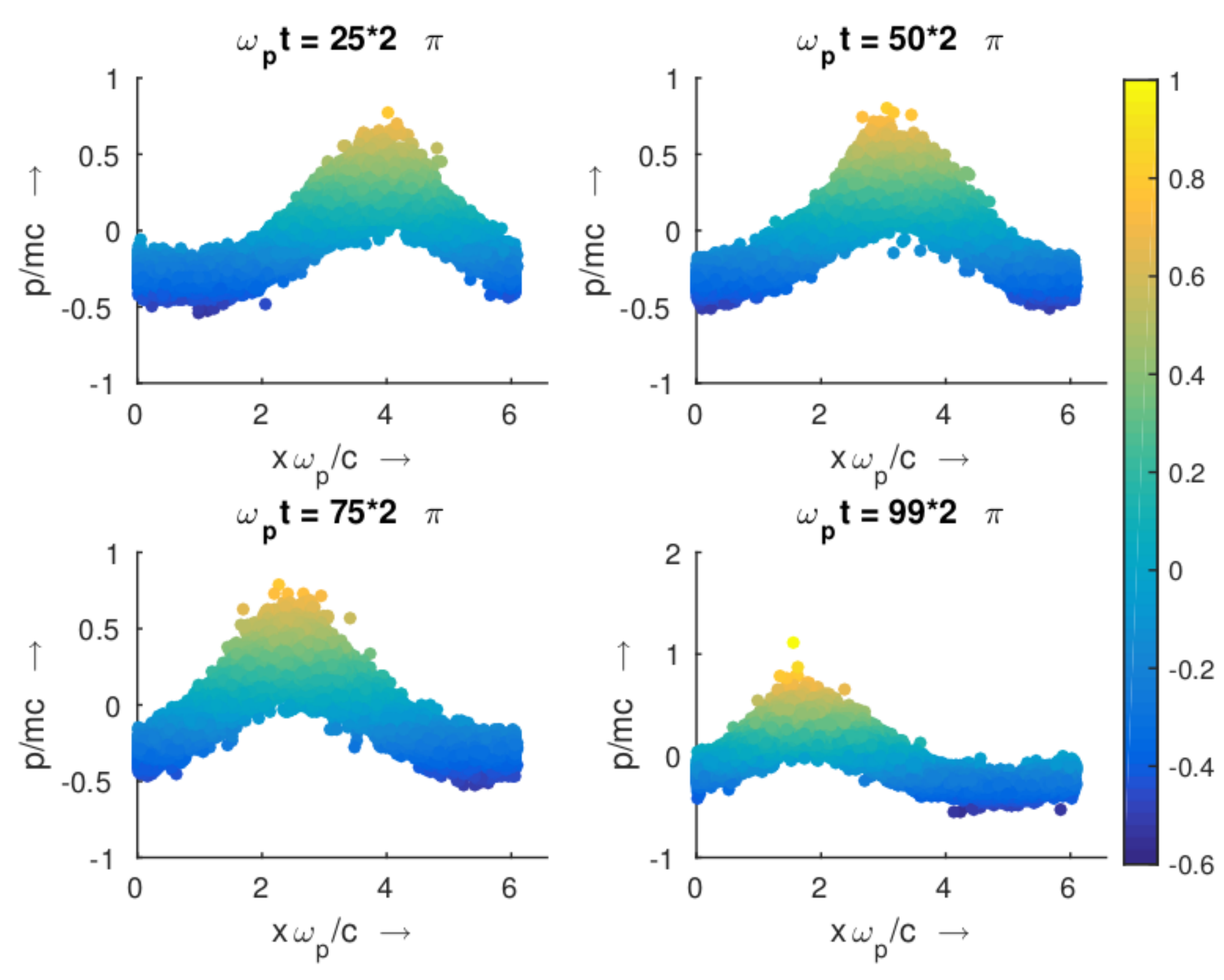}
\caption{Snapshots of phase space for $\beta _{\phi} = 0.95$, $u_m = 0.3$, $\lambda = 0.01$ and $\delta = 0.0$ at different time steps}
\label{fig5j}
\end{figure}

\begin{figure}[htbp]
\centering
\includegraphics[scale=0.24]{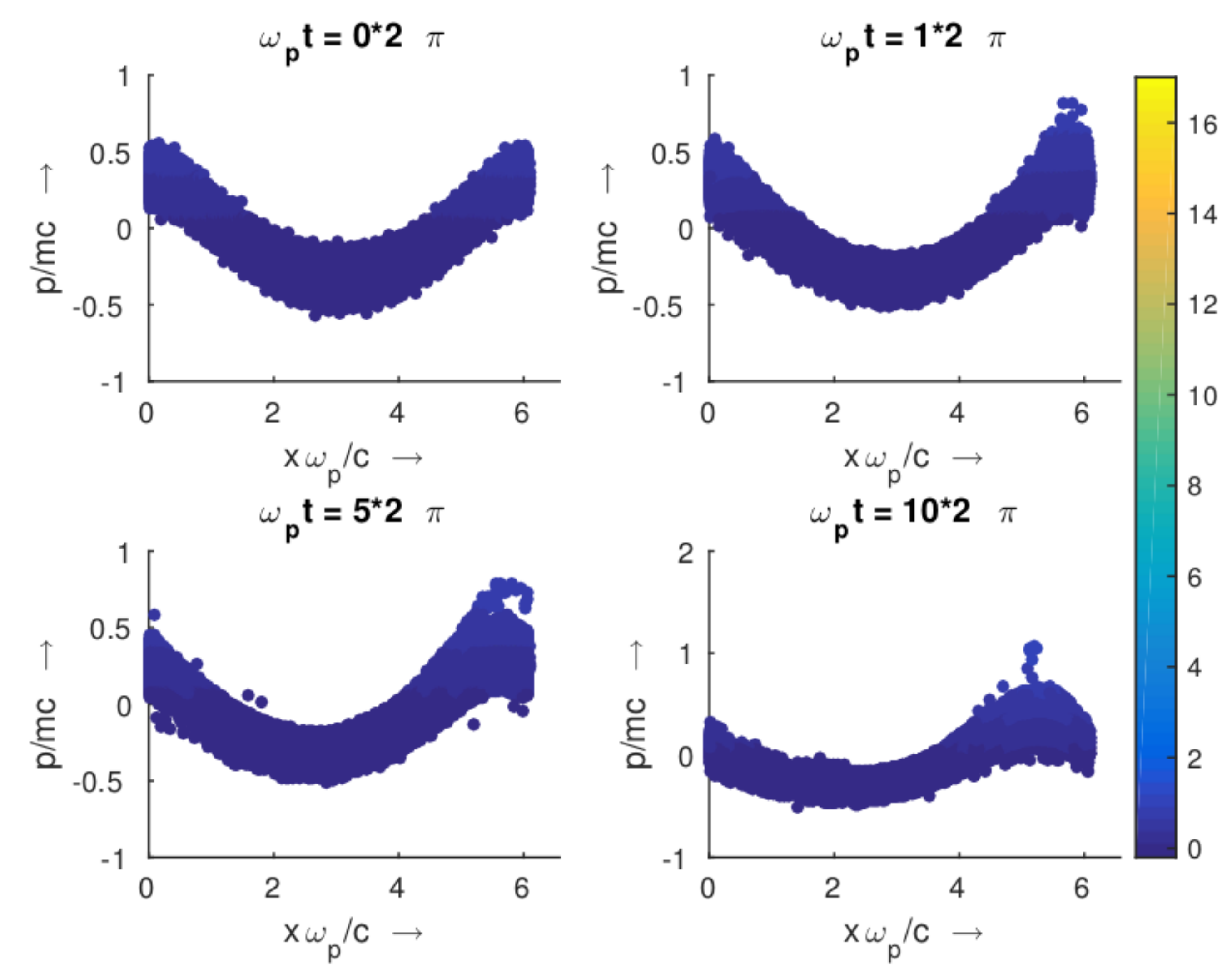}
\includegraphics[scale=0.24]{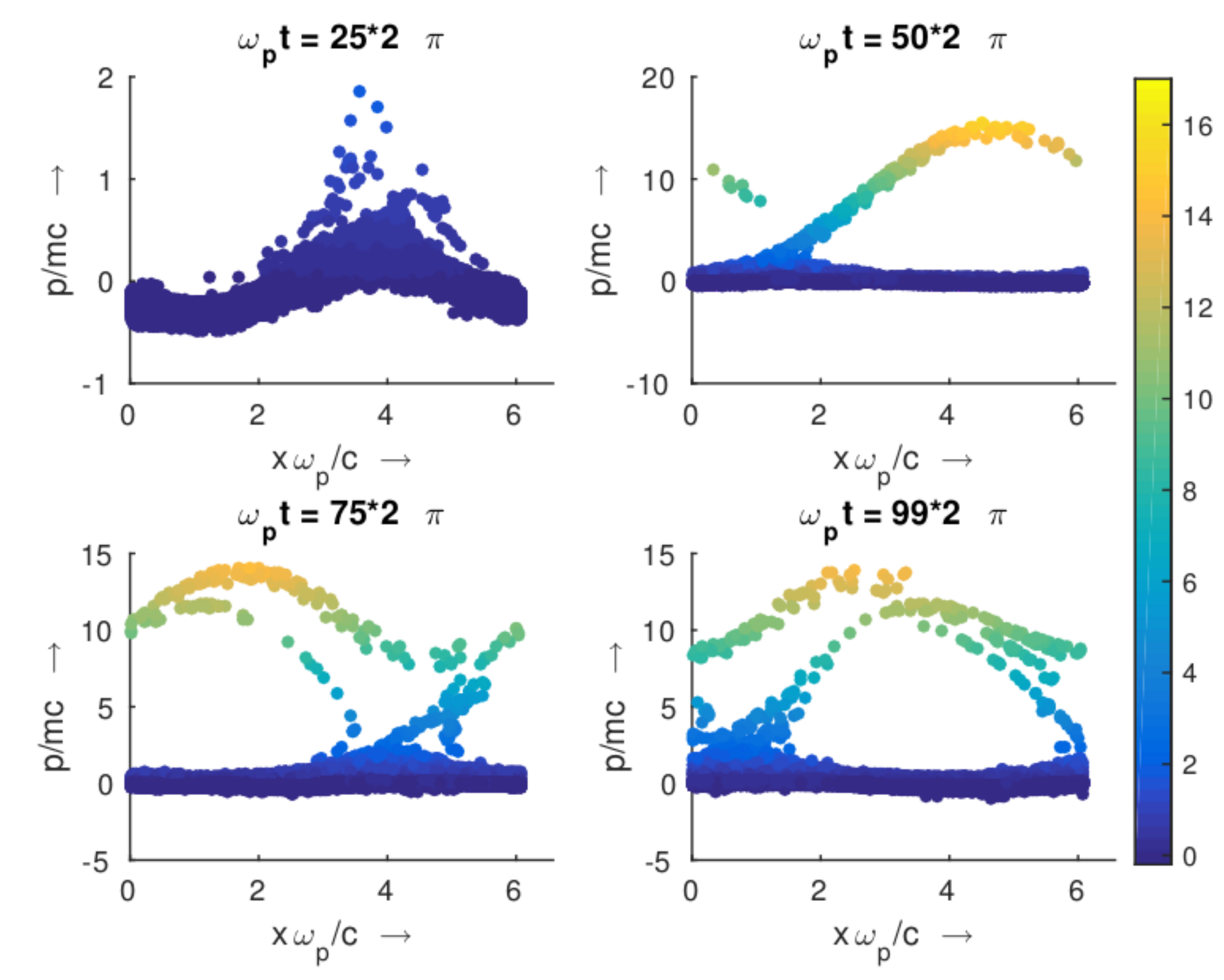}
\caption{Snapshots of phase space for $\beta _{\phi} = 0.95$, $u_m = 0.3$, $\lambda = 0.01$ and $\delta = 0.01$ at different time steps}
\label{fig5k}
\end{figure}

 A clear manifestation of the phase mixing process \textit{i.e} generation of energetic electrons can be seen by plotting the evolution of electron phase - space. In Figs.(\ref{fig5j}) and (\ref{fig5k}) we have respectively plotted the evolution of electron phase space for the cases without and with perturbation. From these figures we observe that in the presence of a small amplitude perturbation, the number of energetic particles increases significantly after wave breaking and thus confirms wave breaking via the gradual process of phase mixing which had been reported earlier by several authors in different contexts \cite{Modena_95, Malka_96, esarey_IEEE, Faure_04, esarey_rev_mod}. We also note that the phase space plots for the unperturbed case remain unchanged from the initial stage of excitation. 

 \section{Estimation of Phase Mixing Time}\label{c5section3}
\begin{figure}[htbp]
\centering
\includegraphics[scale=0.35]{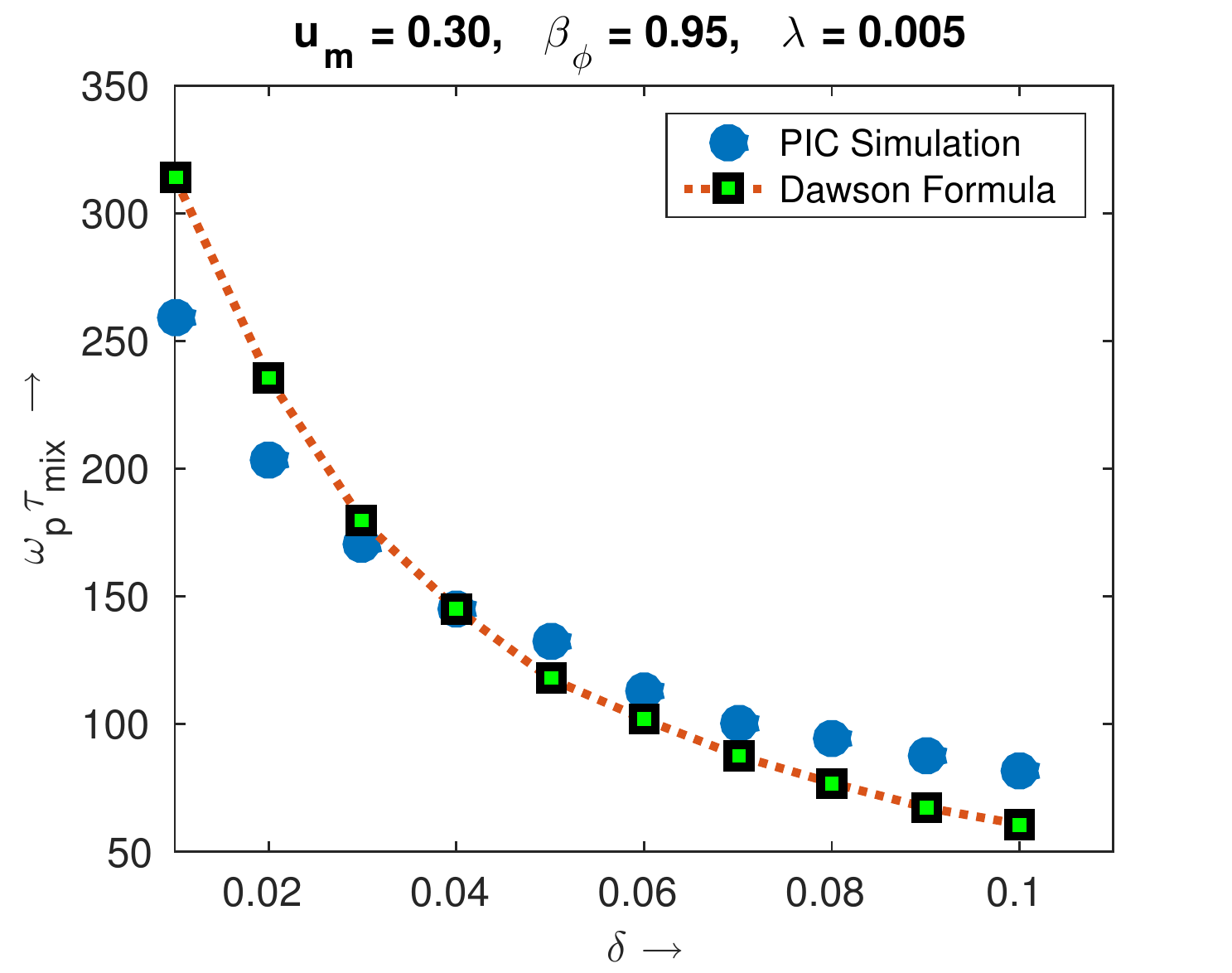}
\includegraphics[scale=0.35]{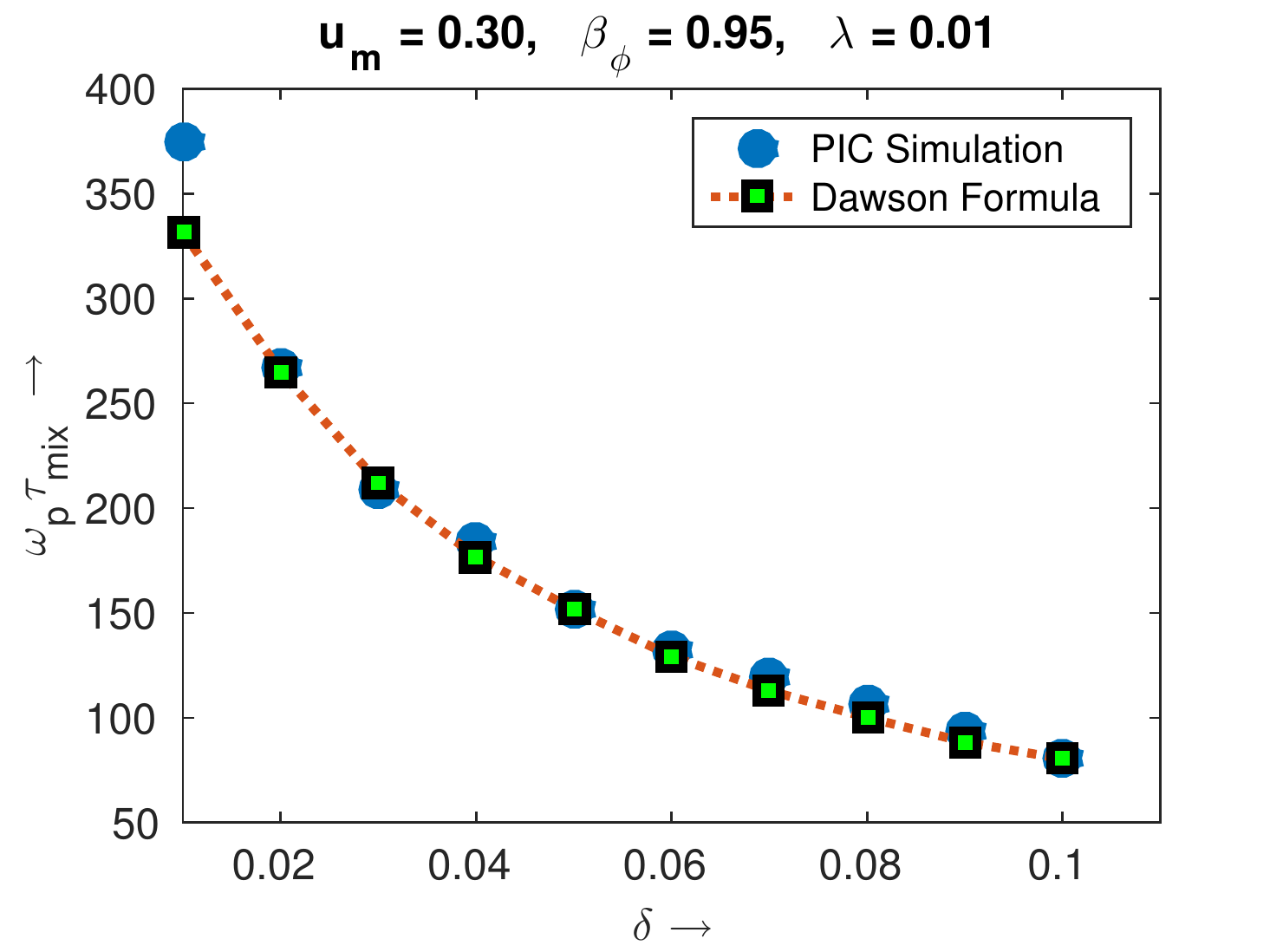}
\caption{Phase mixing time as a function of $\delta$ for $\lambda = 5\times 10^{-3}$ and $10^{-2}$}
\label{fig5n}
\end{figure}

\begin{figure}[htbp]
\centering
\includegraphics[scale=0.35]{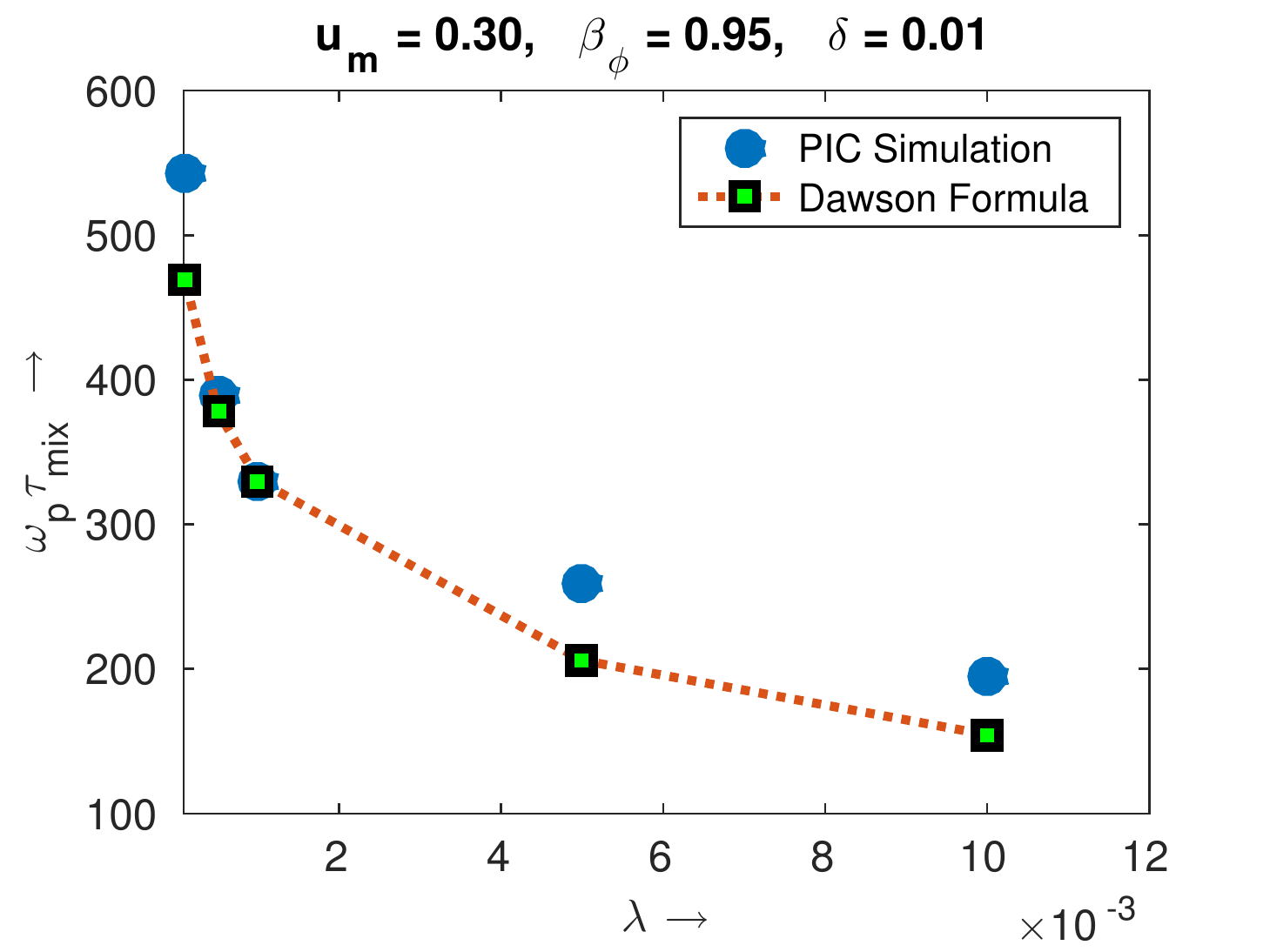}
\includegraphics[scale=0.35]{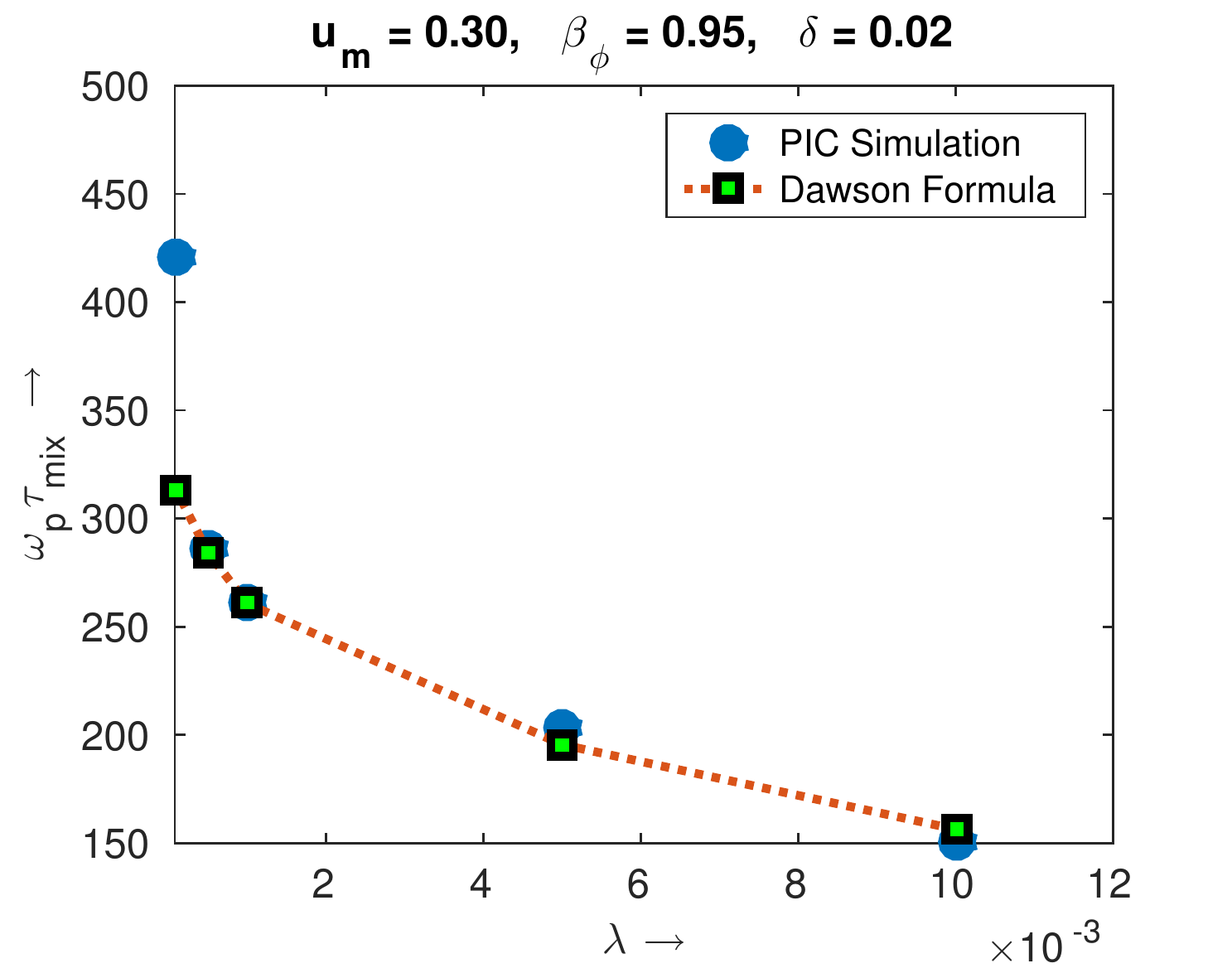}
\caption{Phase mixing time as a function of $\lambda$ for $\delta = 0.01$ and $0.02$}
\label{fig5o}
\end{figure}

As we have found that after adding a small amplitude sinusoidal velocity perturbation the characteristic frequency acquires a spatial dependency, therefore we also expect that in a warm plasma the wave breaking time (phase mixing time) can also be estimated from Dawson's formula \cite{dawson_main} for phase mixing time given for a non-relativistic cold inhomogeneous plasma. This formula was based on out of phase motion of neighbouring sheets constituting the wave and separated by a distance equal to twice the amplitude of the oscillation/wave. Now in order to calculate the phase mixing time scale using Dawson's formula we first measure the derivative $d\Omega /dx$ from Fig.(\ref{fig5y})and note its maximum value. $\xi _{max}$  is calculated by measuring the initial displacement of all the particles from their respective equilibrium positions at $t = 0$. Thus by measuring the values of $d\Omega /dx$ and $\xi _{max}$ from simulation data, we calculate the phase mixing time scale by using Dawson's formula which is $\omega _pt _{mix} \sim \frac{\pi}{2 \xi _{max}d\Omega /dx}$. Now in order to verify this scaling on the amplitude of the perturbation ($d\Omega /dx$ depend on the amplitude of the perturbation $\delta$) we repeat the above numerical experiment such that the maximum velocity amplitude of Akhiezer - Polovin wave is kept fixed at $u_m= 0.30$ and amplitude of the perturbation $\delta$ is varied from $0.01$ to $0.1$. Fig.(\ref{fig5n}) shows the variation of phase mixing time as a function of the amplitude of the applied velocity perturbation $\delta$ for two different values of $\lambda = 5\times 10^{-3}$ and $10^{-2}$. The simulation results clearly indicate that as the amplitude of the perturbation is increased, phase mixing time decreases. Next we vary the electron temperature keeping the values of $u_m$ and $\delta$ fixed. Fig.(\ref{fig5o}) shows the variation of phase mixing time as a function of electron temperature for two different values of fixed $\delta = 0.01$ and $0.02$. This figure indicates that the phase mixing time decreases with increasing the electron temperature $\lambda$. In both two figures blue points are the phase mixing time measured by observing the appearance of first dip of the wave electric field, while the green squares are the phase mixing time scale estimated by using Dawson's formula. These figures show that Dawson's phase mixing time formula clearly captures the underlying physics. The close match between simulation and analytical results (Dawson's phase mixing time formula) supports the role played by phase mixing process in the breaking of relativistically intense electron plasma waves in a warm plasma.
 
\section{Summary and Discussions} \label{c5section4}
In summary, the breaking of a large amplitude relativistically intense electron plasma wave in a warm plasma has been studied by loading a J{\'u}ttner - Synge distribution along with a Akhiezer - Polovin type initial conditions in a relativistic Particle-in-Cell code. It has been observed that in the regime where wave Lorentz factor $\gamma _\phi \ll 1 + \lambda/3$, the wave damps in a few plasma period and swamps out the wave breaking physics. The results obtained from simulations indicate that the damping rate essentially follows the relativistic Landau damping rate predicted by Buti \cite{buti}. In the opposite regime we have found that the wave propagates through the system for a long period of time and in the small amplitude limit, the frequency of the resultant mode follows the relativistic warm plasma dispersion relation $\Omega ^2 = \omega _p^2 + k^2c^2\lambda - 5\omega _p^2\lambda/6$. Next we have shown that when a small amplitude longitudinal sinusoidal velocity perturbation is added to this Akhiezer - Polovin wave, the wave breaks via the process of phase mixing at an amplitude far below its conventional theoretical breaking limits that exist in the literature \cite{Katsouleas_88, rosenzweig_pra_88, scripta, meyer, schroeder_rapid, schroeder_06_POP, trines_2006}. As for example, in Figs.(\ref{fig5i1}) and (\ref{fig5i2}), the wave breaks at an amplitude $eE/m\omega _pc \sim 0.3$, which is far below the wave breaking limits obtained from the models given by Katsouleas-Mori \cite{Katsouleas_88} $\&$ Trines \textit{et. al.} \cite{trines_2006} ($eE/m\omega _pc \sim 1.8033$), Rosenzweig \cite{rosenzweig_pra_88} and Sheng \textit{et. al.} \cite{meyer} ($eE/m\omega _pc \sim 8.1648$) and Schroeder \textit{et. al.} \cite{schroeder_rapid, schroeder_06_POP} ($eE/m\omega _pc \sim 1.6137$). Here we have explicitly shown by carrying out a PIC simulation that, relativistically intense plasma waves in a warm plasma breaks via the process of phase mixing even at $eE/m\omega _pc \sim 0.3$ when perturbed longitudinally by a small amplitude perturbation. Therefore in experiments it is impossible to reach $eE/m\omega _pc \sim 1$, as in a realistic experiment, it is natural to expect some noise which would break the wave via phase mixing. We have also illustrated that this phase mixing time can be predicted from the Dawson's formula \cite{dawson_main} for phase mixing time scale for a non-relativistic cold inhomogeneous plasma, which is based on out of phase motion of neighbouring oscillators constituting the wave and separated by a distance equal to twice the amplitude of the oscillation/wave. 
\bibliographystyle{unsrt}
 \bibliography{arghya_wb}
\end{document}